\documentclass[prd,twocolumn,amsfonts,showpacs,footinbib,preprintnumbers,superscriptaddress]{revtex4-1}

\usepackage{graphicx,bbm,color,amsmath}
\usepackage{dsfont, eufrak}
\let\mathfrak\undefined
\usepackage{amssymb}
\usepackage{amsfonts}
\usepackage{dsfont}
\usepackage{calligra}
\usepackage{calrsfs}
\usepackage{float}
\usepackage[mathscr]{euscript}
\usepackage[utf8]{inputenc} 
\usepackage[colorlinks=true, urlcolor=blue, linkcolor=blue, citecolor=red, hyperindex=true, linktocpage=true]{hyperref}

\newcommand{\ave}[1]{{\langle #1\rangle}}

\newcommand{\be}{\begin{equation}}
\newcommand{\ee}{\end{equation}}
\newcommand{\ii}{ {\rm i} }
\newcommand{\dd}{ {\rm d} }
\newcommand{\ZZ}{\mathbb{Z}}

\def\one{\mathbbm{1}}

\newcommand{\CT}{\mathcal{T}}

\begin{document}

\title{Weak Quantum Chaos}
\author{Ivan Kukuljan}
\affiliation{University of Ljubljana, Faculty of Mathematics and Physics, Jadranska ulica 19, SI-1000 Ljubljana, Slovenia}
\author{Sa\v{s}o Grozdanov}
\affiliation{Instituut-Lorentz for Theoretical Physics, Leiden University,
Niels Bohrweg 2, Leiden 2333 CA, The Netherlands}
\author{Toma\v{z} Prosen}
\affiliation{University of Ljubljana, Faculty of Mathematics and Physics, Jadranska ulica 19, SI-1000 Ljubljana, Slovenia}

\date{\today}

\begin{abstract}
Out-of-time-ordered correlation functions (OTOC's) are presently being extensively debated as quantifiers of dynamical chaos in interacting quantum many-body systems. We argue that in quantum spin and fermionic systems, where all local operators are bounded, an OTOC of local observables is bounded as well and thus its exponential growth is merely transient. As a better measure of quantum chaos in such systems, we propose, and study, the density of the OTOC of extensive sums of local observables, which can exhibit indefinite growth in the thermodynamic limit. We demonstrate this for the kicked quantum Ising model by using large-scale numerical results and an analytic solution in the integrable regime. In a generic case, we observe the growth of the OTOC density to be linear in time. We prove that this density in general, locally interacting, non-integrable quantum spin and fermionic dynamical systems exhibits growth that is at most polynomial in time---a phenomenon, which we term {\em weak quantum chaos}. In the special case of the model being integrable and the observables under consideration quadratic, the OTOC density saturates to a plateau.
\end{abstract}


\maketitle

{\bf Introduction.---}Quantum chaos was an active area of research in the 80's and 90's \cite{GutzwillerBook,HaakeBook,StoeckmannBook}. The main success of the field was a random matrix theory (RMT) classification of universal properties of quantum systems whose classical counterparts are chaotic. The classical limits of such systems have positive Lyapunov exponents, which characterise exponential sensitivity to initial conditions---the so-called {\em butterfly effect}. However, since the (classical) definition of the Lyapunov exponent is based on the concept of phase-space trajectories, one cannot unambiguously translate it to the quantum realm.

Nevertheless, it has been argued that a weaker property of {\em dynamical mixing}---a decay of almost all connected temporal correlators---is sufficient to establish universal quantum chaotic behaviour, such as random matrix statistics of energy spectra \cite{Simons} or the universal exponential decay of Loschmidt echoes \cite{Gorinetal2006}. In the theory of dynamical systems, complex (mixing) dynamics that displays no exponential butterly effect is referred to as {\em weak chaos} (see Ref.~\cite{Klages} and references therein). Examples of such dynamical systems include generic polygonal billiards in which nearby trajectories deviate only linearly with time, while correlation functions nevertheless exhibit mixing \cite{CasatiProsen1999,CasatiProsen2000}.

The study of dynamical mixing (now called {\em scrambling}) and Lyapunov chaos in quantum mechanics was recently revived by the high-energy physics community, initially in the context of the propagation of information in black hole backgrounds \cite{Susskind2008}. In 2014, Kitaev proposed to quantify chaos in interacting quantum many-body systems \cite{Kitaev2014} in terms of the following \textit{out-of-time-ordered} (four-point) \textit{correlation function} (OTOC):
\be
C\left(x,t\right)=-\langle [w_x(t),v_0(0)]^2\rangle_\beta,\label{eq:OTOC}
\ee
where $w_x$, $v_x$ are local observables and $\langle\bullet\rangle_\beta$ denotes the thermal expectation value at inverse temperature $\beta$.
The concept is based on a work by Larkin and Ovchinnikov \cite{Ovchinnikov1969} from 1969, where OTOC was connected to the instability of semi-classical trajectories of electrons scattered by impurities in a superconductor. Consequently, extended quantum systems were defined as {\em chaotic} if there exists a pair of local observables, $w$ and $v$, such that the OTOC \eqref{eq:OTOC} grows exponentially at early times \cite{Ovchinnikov1969,Maldacena2016}:
\be
C\left(x,t\right)\propto e^{\lambda_L \left(t-\left|x\right|/v_B\right)}.\label{eq:ExpGrowth}
\ee
Motivated by the semi-classical picture, $\lambda_L$ is referred to as the \textit{Lyapunov exponent} and $v_B$ the \textit{butterfly velocity}.

A multitude of works examining the properties of quantum chaos have recently been written both from the high-energy perspective (typically in models with long-range interactions and in theories with holographic gravity duals) and from the condensed matter perspective (typically in experimentally more feasible models with local interaction) \cite{Shenker:2013pqa,Roberts:2014isa,Maldacena2016,Polchinski:2016xgd,Maldacena:2016hyu,Jensen:2016pah,Swingle2016,Blake2016,Blake:2016sud,Blake:2016jnn,Lucas:2016yfl,Miyaji2016,Stanford2016,SwingleChowdhury2016,Turiaci:2016cvo,Aleiner:2016eni,Patel:2016wdy,Huang2016,He2016,Fan2016,Chen2016,Hosur2016,Halpern2016,FuSachdev2016,SwingleMeasuring2016,Gaerttner2016,Shen2016,ChenYu2016,Davison:2016ngz,Werner,Rozenbaum,Gritsev}.

In this work, we investigate systems with local interactions with extensive number $N\to\infty$ of degrees of freedom, but with a finite local Hilbert space dimension $D$. In any model with a finite $D$ (including all fermionic and spin lattice models), in which local operators $u$, $v$ are bounded, the exponential growth in (\ref{eq:ExpGrowth}) can be bounded by operator norm inequalities (the triangular inequality, $\| ab\| \le \|a\| \|b\|$ and $\ave{a}_\beta \le \| a\|$):
\be
C\left(x,t\right) \le 4 \left\| v \right\|^2 \left\| w \right\|^2.\label{eq:Bound}
\ee
Thus, the OTOC can only grow exponentially up to a finite (scrambling) time $t^*$, after which it remains bounded by a constant. This is consistent with the observations made in other works on OTOC's (of local observables) in fermionic systems where OTOC's were always observed to reach a plateau \cite{Aleiner:2016eni,Huang2016,He2016,Fan2016,Chen2016,Hosur2016}. As already noted in \cite{Patel:2016wdy}, the only way for the exponential time evolution to persist to late times is if there is a small prefactor multiplying the exponential function in \eqref{eq:ExpGrowth}. Even in the Sachdev-Ye-Kitaev (SYK) model with long-range interactions, this prefactor is $1/N$, which becomes small as $N\to\infty$ \cite{Polchinski:2016xgd}. Exponential growth \eqref{eq:ExpGrowth} of the OTOC is therefore at best a transient effect in systems of interest to this work.

If interactions are local, $C(x,t)$ can be further bounded by the Lieb-Robinson theorem (LRT) \cite{LiebRobinson1972} (see also \cite{Swingle2016}):
\be
C\left(x,t\right) \le 4 \left\| v \right\|^2 \left\| w \right\|^2 e^{-\mu \max\{0,\left|x\right|-v_{\!L\!R}t\}}.
\ee
In this case, for $t\ll t^* = \left|x\right|/v_{\!L\!R}$, the OTOC is even more suppressed. The interpretation of this effect is clear: namely, $t^*$ is the time in which $C(x,t)$ enters the causal cone. Before $t^*$, $C(x,t)$ is almost zero, while after $t^*$, it is bounded by \eqref{eq:Bound} and saturates at a plateau. The dynamics can only be non-trivial near the edge of the causal-cone (or for $t \sim t^*$), where $C(x,t)$ can vary greatly. This is consistent with \cite{Aleiner:2016eni,Huang2016}.

Another important fact is that momentum operators---the observables that Ref.~\cite{Ovchinnikov1969} originally used to compute the Lyapunov exponent of the semiclassical trajectories---are unbounded. Therefore, if we wanted to preserve the semiclassical justification of the OTOC, which is necessary to be able to speak about quantum chaos, the quantum observables under consideration must have unbounded spectra.

These observations can be summarised in the intuitive statement that if chaos is to fully develop over long time, the observables have to provide enough ``space" for this to happen; they need to be unbounded. Indeed, this is the case with general observables in bosonic systems (usually studied in holography). However, this condition is not fulfilled by local observables in fermionic or spin systems, or more generally, in systems with a finite $D$. On the other hand, extensive observables in such theories do satisfy the unbounded spectrum criterium and therefore have the capacity to fully unveil the system's dynamical properties and quantum chaos. Motivated by this fact, we propose a new measure of quantum chaos: the {\em density of the OTOC} (dOTOC) of (non-local) extensive operators $V \equiv \sum_{x\in\Lambda} v_x$, $W \equiv \sum_{x\in\Lambda} w_x$, with $w_x,v_x$ local. It is defined on a $d-$dimensional lattice $\Lambda$ with $N$ sites as the centralised second moment of the commutator
\be
c^{(N)}(t) := -\frac{1}{N}\left( \ave{[W(t),V(0)]^2}_\beta-\ave{[W(t),V(0)]}^2_\beta\right).
\label{eq:dOTOC}
\ee
The disconnected part, which is just the square of the standard dynamical susceptibility (i.e. the response function), has been subtracted to make the dOTOC well defined in the thermodynamic limit (TL) for any temperature. Because of the cyclicity of the trace, this term vanishes at $\beta=0$ (this will occur in the model that we study below). Using the LRT and the clustering property of thermal states, which holds for any temperature in $d=1$ \cite{Araki69} and for sufficiently high temperature in $d>1$ \cite{Eisert}, in Appendix \ref{proofbound}, we rigorously prove that the dOTOC satisfies a uniform (in $N$) polynomial bound
\be
c^{(N)}(t) \le A t^{3d},
\label{eq:polbound}
\ee
where $A$ is an $(N,t)-$independent constant. The same bound equally holds in the TL, $c(t):=\lim_{N\to\infty} c^{(N)}(t)$.

Moreover, we report below the results of extensive numerical and analytical calculations, which demonstrate that possibly the simplest non-trivial locally interacting quantum chaotic spin system: the kicked Ising (KI) quantum spin chain \cite{ProsenPTPS2000,ProsenPRE2002}, exhibits linear growth of the dOTOC of extensive magnetisation observables, $c(t) \propto t $. An exception is the integrable KI model (equivalent to a free fermion model), for which we show analytically that its dOTOC of extensive quadratic observables (in fermionic variables) saturates, $c(t\to\infty) = {\rm const}$. Since the KI model seems to be generic, we further conjecture that the bound (\ref{eq:polbound}) is not optimal and that typical one-dimensional, non-integrable and locally interacting models exhibit linear growth of dOTOC's. 

As a consequence, theories under consideration in this work are not expected to exhibit any late-time butterfly effect, but as we know from results in the RMT, can still be chaotic. In reference to classical mixing systems without the butterfly effect, we term the phenomenon of infinite polynomial growth of dOTOC's {\em weak quantum chaos}. 

{\bf Kicked quantum Ising model.---}The Hamiltonian of the one-dimensional KI model consists of the Ising-interaction term $H_\text{Ising}=\sum_{j}J\sigma_{j}^{x}\sigma_{j+1}^{x}$ and the kick term $H_\text{kick}=\sum_{j}h\left(\sigma_{j}^{z}\cos\varphi+\sigma_{j}^{x}\sin\varphi\right)$:
\begin{align}
H (t) = H_\text{Ising}+H_\text{kick}\sum_{n\in\mathbb{Z}}\delta\left(t-n\right), \label{eq:HamiltonianKI}
\end{align}
where $\sigma^{\alpha}_j$ are local Pauli spin operators. The model has three parameters: the Ising coupling $J$, the magnitude of the external magnetic field $h$ and the inclination of the external magnetic field $\varphi$. KI is a periodic (in time) system with the Floquet propagator:
\begin{eqnarray}
U&=&\CT \left\lbrace e^{-\ii \int_0^1\dd t H(t)}\right\rbrace \nonumber\\
&=&e^{-\ii J\sum_j\sigma_{j}^x\sigma_{j+1}^x}e^{-\ii h\sum_j\left(\sigma_{j}^{z}\cos\varphi+\sigma_{j}^{x}\sin\varphi\right)}. \label{eq:Floquet}
\end{eqnarray}
Because of the temporal periodicity, KI dynamics can be viewed as discrete in time, or as a quantum cellular automaton. The effect of a perturbation on a single lattice site propagates in a causal-cone with speed $1$. Namely, information can spread only by one site, left or right, within one period (kick of the magnetic field). Random matrix analysis \cite{PinedaProsenPRE2007,KukuljanProsen2016} revealed that KI is chaotic.

The system has a further nice property of being integrable (quasi-free) for transverse magnetic field, $\varphi=0$, and non-integrable (and interacting) for $\varphi>0$. Thus, $\varphi$ serves as a handy parameter which allows us to study integrability breaking. See e.g. \cite{ProsenPRE2002,ProsenJPA2007} for a survey of elementary dynamical properties of the KI model.

Here, we study the KI chain with $N$ spins and evaluate the dOTOC (\ref{eq:dOTOC}) $c^{(N)}_\alpha(t)$ for a 
(non-local) extensive magnetisation, $W=V=M_\alpha = \sum_{j=1}^N\sigma_j^\alpha$,
which can either be transverse ($\alpha=z$) or parallel ($\alpha=x$) to the direction of the Ising interaction. We take $\beta=0$ as an infinite-temperature Gibbs ensemble is the only meaningful equilibrium state for periodically driven systems, which generically heat up to infinite temperature. We use three different approaches, two numerical methods for the general inclination ($0\leq\varphi\leq\frac{\pi}{2}$) and an analytical solution for the transverse field case $\varphi=0$. In the first, appropriate for small system sizes (up to $N\sim12$), we used the exact numerical Floquet operator \eqref{eq:Floquet}. The second method, used for intermediate system sizes (up to $N\sim22$), was a Monte-Carlo wave-function sampling based on typicality arguments (explained in Appendix \ref{app:Typicality}). The analytical solution in the TL  for the integrable (transverse) case and transverse magnetisation $M_z$, was found using fermionisation. We outline the main steps for obtaining the analytical solution in what is to follow. 

{\bf Analytical solution.---}For the transverse field ($\varphi=0$), KI is a quasi-free model. If, furthermore, the (extensive) observable of interest is simple enough, the dOTOC allows for an analytic solution in terms of Jordan-Wigner transformation of Pauli spins into staggered Majorana fermion operators 
\begin{align}
w_{2j}=\left(\prod_{k<j}\sigma_k^z\right)\sigma_j^x, && w_{2j+1}=\left(\prod_{k<j}\sigma_k^z\right)\sigma_j^y,
\end{align}
obeying the anti-commutation relations $\left\{w_i,w_j\right\}=2\delta_{ij}$. The Floquet operator \eqref{eq:Floquet} then takes the following form:
\be
U=e^{-J\sum_jw_{2j-1}w_{2j}}e^{-h\sum_jw_{2j}w_{2j+1}} . \label{eq:FloquetFermion}
\ee
It is clear from \eqref{eq:FloquetFermion} that the KI model is free for $\varphi = 0$ \footnote{We note that even free theories can exhibit complicated entangled collective behaviour when one considers the dynamics of composite operators. See e.g. \cite{Grozdanov:2015nea,Polonyi:2015yia}.}. Now, the transverse magnetisation can be expressed as a sum of quadratic Majorana operators:
\begin{align}
M_z=-\ii\sum_{j\in\ZZ}w_{2j}w_{2j+1}, \label{eq:MzMajorana}
\end{align}
which enables the analytic computation of the dOTOC of $M_z$ \footnote{Longitudinal magnetisation is instead a sum of infinite strings of fermions, $M_x=\sum_j\prod_{k<j}\left(-\ii w_{2k}w_{2k+1}\right)w_{2j}$. Computation of the OTOC for $M_x$ is therefore significantly more involved.}. Power-expanding the Floquet operator \eqref{eq:FloquetFermion} and using $\left(w_i w_j\right)^2=-1$ for $i\neq j$, $U$ further simplifies to 
\begin{align}
U&=\prod_j\left(\cos\left(J\right)-w_{2j-1}w_{2j}\sin\left(J\right)\right)\cdot\nonumber\\
& \cdot   \prod_k\left(\cos\left(h\right)-w_{2k}w_{2k+1}\sin\left(h\right)\right)   =   U_{\text{Ising}}U_{\text{kick}} \, .\label{eq:FloquetMajorana}
\end{align}

Since the transverse field model is free, it is convenient to work in the Fourier transformed Majorana basis: 
\begin{align}
w(\theta)=\sum_{j}w_{2j}e^{\ii \theta j}, && w'(\theta)=\sum_{j}w_{2j+1}e^{\ii \theta j},
\end{align}
with shorthand notation
$
\underline{w}(\theta)=
\begin{pmatrix}
	w(\theta)\\
	w'(\theta)
\end{pmatrix}.
$
One can show (Appendix \ref{app:KIFourier}) that the Floquet propagator in the Heisenberg picture,
$
\mathcal{U}\underline{w}(\theta):=
\begin{pmatrix}
	U^\dagger w(\theta) U\\
	U^\dagger w'(\theta) U
\end{pmatrix}
$,
takes the following form in Fourier transformed Majorana basis:
\begin{align}
&\mathcal{U}(J,h,\theta)=\mathcal{U}_{\text{kick}}(J,h,\theta)\mathcal{U}_{\text{Ising}}(J,h,\theta)     \label{eq:KIFloquetFourier}\\
&= \left(
\begin{matrix}
\cos(2h) & -\sin(2h)\\
\sin(2h) & \cos(2h)
\end{matrix} \right) 
\left(
\begin{matrix}
\cos(2J) & e^{\ii \theta}\sin(2J)\\
-e^{-\ii \theta}\sin(2J) & \cos(2J)
\end{matrix} \right).\nonumber    
\end{align}
This $2\times 2$ unitary matrix valued symbol can be diagonalised as:
\be
\mathcal{U}(J,h,\theta)=V^\dagger(J,h,\theta)\left(\begin{matrix}
	e^{\ii \kappa(J,h,\theta)}& \\ & \!\!\!\!\! e^{-\ii \kappa(J,h,\theta)}
\end{matrix}\right) V(J,h,\theta),\label{eq:FloquetEigendecomp}
\ee
where
\begin{align}
\kappa(J,h,\theta)=& \, \arccos\left[\cos(2J)\cos(2h)\right.  +  \nonumber\\
&+\left.\cos(\theta)\sin(2J)\sin(2h)\right] ,  \label{eq:Spectrum}
\end{align}
and $V(J,h,\theta)$ is given explicitly in Appendix \ref{app:SpectrumKI}.

Knowing that the KI Majorana fermions in the Fourier basis time evolve as
$
\underline{w}(\theta,t)=\mathcal{U}(\theta)^t\underline{w}(\theta,0)
$
allows us to define the real space propagator as:
\be
K_{ab}^{kj}(t):=\left\langle w_{2k+a-1}\,w_{2j+b-1}(t)\right\rangle, \label{eq:NotationSymbol}
\ee
for $a,b\in\left\lbrace1,2\right\rbrace$. $K^{kj}$ can then be computed from $\mathcal{U} (\theta)$ (Appendix \ref{app:Symbol}):
\be
K^{kj}(t) := K^{j-k}(t)=\frac{1}{2\pi}\int_{-\pi}^{\pi}\dd\theta e^{-\ii \theta \left(j-k\right)} \mathcal{U}^t(\theta).\label{eq:SymbolFinal}
\ee

Using the propagator \eqref{eq:SymbolFinal}, we can compute the infinite temperature OTOC of the transverse magnetisation, $c^{(N)}_z(t)$. First, we express the terms in \eqref{eq:dOTOC} using \eqref{eq:MzMajorana}, e.g. $\ave{\sigma_i^z (t) \sigma_j^z \sigma_k^z (t ) \sigma_l^z}$ as an eight-fermion expectation value $\ave{w_{2i} (t) w_{2i+1} (t) w_{2j}w_{2j+1}w_{2k} (t) w_{2k+1} (t) w_{2l} w_{2l+1}}$. Then, using $\eqref{eq:NotationSymbol}$, these are expressed as the product of four propagators (one for each time-dependent fermion) times an equal-time eight-fermion expectation value, with terms summed over four spatial and spin indices 
(see Appendix \ref{app:FormulaOTOC} for details). Simple algebraic manipulations then lead to the final expression for the dOTOC in the TL:

\begin{widetext}
\begin{eqnarray}
c_z(t)&=&-4\sum_{j,l_1,l_3\in\ZZ}^{j\neq 0} \sum_{s_0,s_j,p_1,p_3=1}^2
\left(-1\right)^{p_1+p_3}K_{S(p_1),1}^{R_1(p_1)}(t)K_{S(\tilde{p_1}),2}^{R_1(\tilde{p_1})}(t)\cdot\nonumber\\
&&
\hspace{3cm}\cdot\left[\left(-1\right)^{s_j+s_0}K_{\tilde{S}(p_3),1}^{R_3(p_3)}(t)K_{\tilde{S}(\tilde{p_3}),2}^{R_3(\tilde{p_3})}(t)
-
K_{S(p_3),1}^{R_3(p_3)}(t)K_{S(\tilde{p_3}),2}^{R_3(\tilde{p_3})}(t)\right],\label{eq:OTOCsimplified}
\end{eqnarray}
\end{widetext}
where we used the following notation:
$
R_1:=(
	l_1-j,
	l_1)$,
$R_3:=(
	l_3-j,
	l_3)$,
$S:=(s_j,
	s_0),
$
together with the notation $v=(v(1),v(2))$ for vector components and $\tilde{1}:=2$, $\tilde{2}:=1$. 
We can use the formula \eqref{eq:OTOCsimplified} in two different ways. For intermediate times $t\sim50$, we can perform the integral in \eqref{eq:SymbolFinal} exactly and evaluate the sums in \eqref{eq:OTOCsimplified}, which, because of the causal-cone spreading of information, now become finite sums (see Appendix \ref{app:NumaricalApplication} for details).

Secondly, we can use the stationary phase approximation in combined \eqref{eq:SymbolFinal}, \eqref{eq:FloquetEigendecomp} and \eqref{eq:OTOCsimplified} to compute the large-$t$ asymptotic behaviour of the dOTOC. In this way, we prove that for large times, $c_z(t)$ is a constant (dependent only on $J$ and $h$). In other words, the dOTOC of quadratic extensive observables in the integrable KI model saturates to a plateau. Details are explained in Appendix \ref{app:Asymptotics}.

\begin{figure*}
	\includegraphics[width=0.9\linewidth]{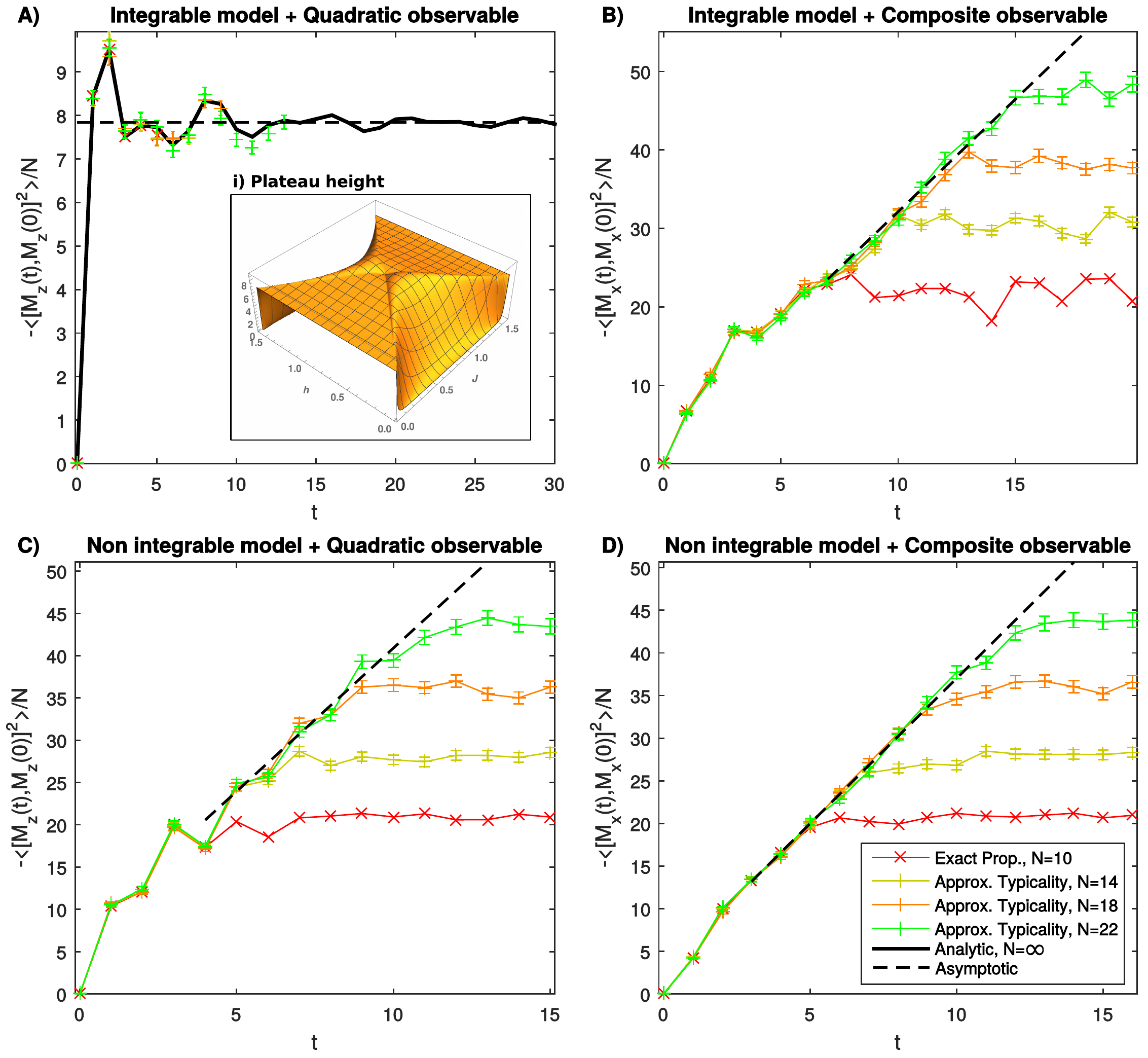}
	\caption{\label{fig:OTOC} Density of the OTOC of extensive observables for one-dimensional KI model \eqref{eq:HamiltonianKI} with periodic boundary conditions is presented for four possible regimes. In the upper panels (A, B), the magnetic field is transversal ($\varphi=0$) so the system is integrable (free), while in the lower panels (C, D) the field is tilted ($\varphi=\frac{\pi}{4}$) so the model is non-integrable. In the left panels  (A, C) the observable is a sum of quadratic Majorana terms \eqref{eq:MzMajorana}, while in the right panels (B, D), the observable is a sum of terms composed of infinite Majorana strings  (composite). 
Here $J=0.7$ and $h=1.1$ but the behaviour was found qualitatively similar for other values of $J,h$. The numerically exact results for small system sizes are plotted with crosses. Results obtained with numerical method based on typicality arguments (with a sample of $50\times50$ random vectors) are plotted with error bars. The analytical solution for the integrable case and quadratic observable is plotted with a bold black line. The asymptotic behaviour in the limits $N\rightarrow\infty$ and $t\rightarrow\infty$ is plotted with a dashed line. In the integrable + quadratic case the dashed line is the result of our analytic solution. In other cases it is an extrapolation based on numerics. The numerical results start to deviate around $t\sim N/2$ due to finite size effects. The inset (i) shows the dependence of the plateau height on the parameters $J$ and $h$.}		
\end{figure*}

{\bf Results and discussion.---}In summary, we observe two distinct behaviours of the OTOC density for extensive observables in a one-dimensional KI model. For a generic situation, unless the model is integrable and the observable quadratic, the extensive dOTOC grows linearly with time. In fact, numerical results for finite system sizes saturate to a plateau at $t\sim N/2$, but this is simply due to a finite size effect---a consequence of the causal cone coming around the periodic boundary. This plateau grows with an increasing system size $N$ and we expect that it disappears in the TL $N\rightarrow\infty$. In the regime where the model is integrable (free) and the observable is simple (quadratic in fermion operators), the dOTOC saturates to a genuine plateau despite the fact that the spectrum of the observable is unbounded. The latter statement was proven in this work by finding an explicit analytic solution for $c_z(t)$ from which the expression for the height of the plateau for a given set of parameters $J$ and $h$ could be found. The results of the time dependence of the extensive dOTOC for different scenarios are presented in Figure \ref{fig:OTOC} and explained in the caption.

For the integrable case with $\varphi=0$, the quasiparticle spectral gap closes on the line of $J=h$ in the parameter space and the system exhibits a Floquet analogue of a quantum phase transition, i.e. $\kappa\left(J=h,\theta=0\right)=-\kappa\left(J=h,\theta=0\right)=0$ (cf. \eqref{eq:FloquetEigendecomp}, \eqref{eq:Spectrum}). It is interesting to ask whether the OTOC also reflects this transition in any way. What we find is that the plateau height  ceases to be smooth for $J=h$. Beyond that, we also checked the slope of the OTOC for longitudinal magnetisation $M_x$ in the vicinity of this line. It turns out that the slope exhibits a peak, but not exactly on the line $J=h$. This could be the effect of a small system size, which was necessary for numerics. It is plausible that the peak may align with $J=h$ in the TL. 

This work should be considered as a starting point for future investigations of quantum, weakly chaotic systems, which exhibit dynamical late-time mixing but do not display any exponential butterfly effect due locality of interactions and finiteness of the local Hilbert space. In such systems, the standard OTOC rapidly plateaus and is therefore not a good measure of chaos. This observation led us to propose of a new measure of chaos: density of the OTOC of non-local extensive operators. We have proven (Appendix \ref{proofbound}) that such correlators always exhibit a polynomial bound and can thus be widely used to diagnose and classify quantum chaos. In the case of the non-integrable KI model studied here, the growth is linear. Intuitively, it seems apparent that in locally interacting systems, information propagates slower than in an all-to-all interacting theory like the SYK model.
The speed is limited by the Lieb-Robinson velocity. However, what is less apparent is that such systems can still be chaotic; a result established by an RMT analysis \cite{PinedaProsenPRE2007}.   

Lastly, we note that in order to study chaos in strongly coupled, large-$N$ theories (even in those that do exhibit the buttery effect), it would be interesting to extend holographic calculations to computations of OTOC's of non-local, smeared operators. For detailed future analyses, we will likely need to utilise the full machinery of holographic $n$-point function calculations \cite{Barnes:2010jp,Arnold:2011hp,Grozdanov:2014kva} that will extend beyond studying gravitational shock waves \cite{Shenker:2013pqa,Roberts:2014isa}.   

{\bf Acknowledgements.---} We thank M. Medenjak for fruitful discussions and E. Ilievski  for useful comments. The work has been supported by the ERC grant OMNES and grants P1-0025, P1-0044 of Slovenian Research Agency (ARRS). S. G. is supported in part by a VICI grant of the Netherlands Organization for Scientific Research (NWO) and by the Netherlands Organization for Scientific Research/Ministry of Science and Education (NWO/OCW).

\bibliography{KGP_OTOC}

\onecolumngrid

\newpage

\appendix
\section{Proof of the polynomial bound on the density of extensive OTOC}
\label{proofbound}
In this section, we first prove that the density of the extensive OTOC (dOTOC) for 1D locally interacting translationally invariant lattice systems with finite local Hilbert space dimension cannot grow faster than with the third power of time. Then, we directly extend our theorem, Eq. (\ref{eq:polbound}) of the paper, to $d-$dimensional regular lattices.

To derive the bound, we will take advantage of two important theorems that hold for locally interacting lattice systems. The first, the \textit{Lieb-Robinson theorem} (LRT) \cite{LiebRobinson1972}
states that for any locally interacting lattice system there exist positive constants $\xi$, $\mu$  and $v_{LR}$, such that for any two operators $a$ and $b$:
\be
\left\|\left[a(t),b\right]\right\|\leq \xi \min\left\{\left|\text{supp}(a)\right|,\left|\text{supp}(b)\right|\right\}  \left\|a\right\|\left\|b\right\|e^{-\mu \max\left\{0,\dd\left(\text{supp}(a),\text{supp}(b)\right)-v_{\!L\!R}t\right\}}.\label{eq:LiebRobinson1App}
\ee
Here, $\text{supp}(a)\subset\ZZ$ denotes the support of a local operator $a$ and $\dd\left(\bullet,\bullet\right)$ is the distance between two sets. Roughly speaking, the theorem says that the commutator of two local observables grows in a causal-cone, spreading with velocity $v_{\!L\!R}$. 

The authors of Ref.~\cite{BHV06} have found an elegant and useful reformulation of the LRT. Let $\Gamma$ be a subset of the lattice of $N$ sites and define 
\be
\left.a\right|_\Gamma := \frac{\text{tr}_{\Gamma^C} a}{\text{tr}\one_{\Gamma^C}}\otimes \one_{\Gamma^C},
\ee
where $\Gamma^C$ denotes the set complement,  to be a projection of the operator $a$ on the sublattice $\Gamma$. Note that $\text{supp}(a|_{\Gamma}) = \Gamma$. Then, for a given locally interacting system, the LRT is equivalent to \cite{BHV06,IP13}
\be
\left\|a(t)-\left.a(t)\right|_\Gamma\right\|\leq \xi \, \left|\text{supp}(a)\right| \left\|a\right\|e^{-\mu\max\left\{0,\dd\left(\text{supp}(a),\Gamma^C\right)-v_{\!L\!R}t\right\}}.\label{eq:LiebRobinson2App}
\ee

The second theorem that we will need is the \textit{exponential clustering property} of thermal states \cite{Araki69,Matsui2003}. For a thermal state of a one-dimensional locally interacting
system, there exist positive constants $\chi$ and $\rho$, such that the following inequality is satisfied by any two operators $a$ and $b$:
\be
\left|\ave{a,b}_\beta^{c}\right|\leq \chi \left\|a\right\|\left\|b\right\|e^{-\rho\,\dd\left(\text{supp}(a),\text{supp}(b)\right)},\label{eq:ClusteringApp}
\ee
where, in order to make the expressions in this section more compact, we have introduced the notation for the \textit{connected (bipartite) correlation function}:
\be
\ave{a,b}_\beta^{c}:=\ave{ab}_\beta-\ave{a}_\beta\ave{b}_\beta.
\ee
An analogous result is true for locally interacting Hamiltonians on arbitrary $d-$dimensional lattices for sufficiently high temperatures \cite{Eisert}. As we will show, the three bounds, \eqref{eq:LiebRobinson1App},  \eqref{eq:LiebRobinson2App} and \eqref{eq:ClusteringApp}, imply a polynomial bound for the dOTOC.

Our goal is to compute an upper bound on the dOTOC:
\begin{eqnarray}
c(t)&:=&\lim_{N\rightarrow\infty}c^{\left(N\right)}(t)\nonumber\\
&:=&-\lim_{N\rightarrow\infty}\frac{1}{N}\sum_{i,j,k,l\in\ZZ}\left\langle \left[w_{i}(t),v_{j}\right],\left[w_{k}(t),v_{l}\right]\right\rangle_\beta^c\nonumber\\
&=&-\sum_{i,j,k\in\ZZ}\left\langle \left[w_{i}(t),v_{j}\right],\left[w_{k}(t),v_{0}\right]\right\rangle_\beta^c\nonumber\\
&=:&\sum_{i,j,k\in\ZZ}c_{ijk}(t).\label{eq:ExtensiveOTOCapp}
\end{eqnarray}
Note that in the third line, we used the translational invariance of the system.

\begin{figure*}
	\includegraphics[width=0.65\linewidth]{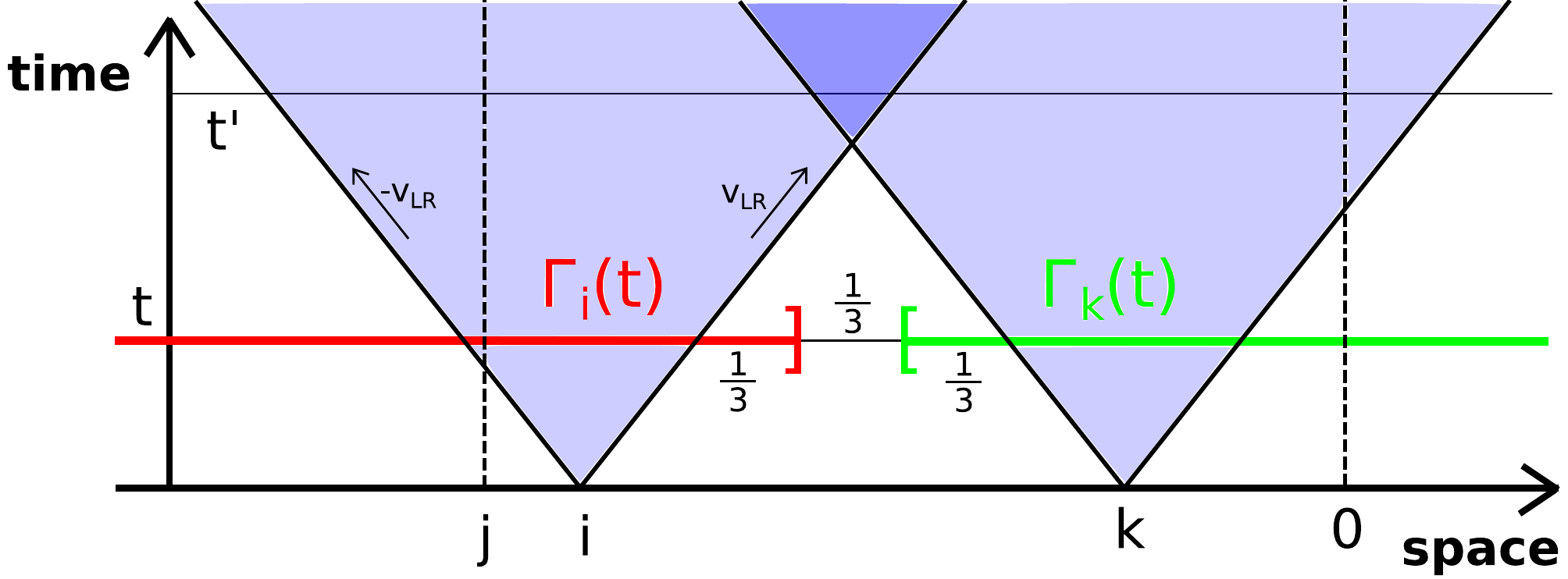}
	\caption{\label{fig:Bound} The illustration of the main concepts needed in proving the polynomial upper bound on the dOTOC. At a given time, we can divide the $i$--$k$ plane into two regions and use different techniques to bound the contribution to the total bound on OTOC coming from each region.  We will use the intuition implied by the LRT \eqref{eq:LiebRobinson1App} that a commutator spreads essentially in causal-cone and is exponentially damped outside. The first region is the one where the causal-cones corresponding to the two commutators in \eqref{eq:ExtensiveOTOCapp} overlap (for the particular choice of $i$ and $k$ in the drawing, this is the case for example at time $t'$). The leading order term in the bound on OTOC ($\propto t^3$) will come from this region. The second is the region where the light cones are well separated and can be embedded into semi-infinite intervals ($\Gamma_i$, $\Gamma_k$) with growing distance between them. This region will contribute subleading terms ($\propto t^2$). }		
\end{figure*}

The bound can be established by first using the triangular inequality
\be
c(t)\leq\sum_{i,j,k\in\ZZ}\left|c_{ijk}(t)\right|,
\ee
and then by finding the appropriate bounds for individual terms. To take advantage of the exponential clustering property, at every time $t$, we will separate the $i$--$k$ plane into two domains. The first, $\left|k-i\right|\leq2v_{\!L\!R}t$, is the region where the causal-cones of the two commutators in \eqref{eq:ExtensiveOTOCapp} are overlapping. There, the exponential clustering cannot be used, but the region is bounded in $\left|k-i\right|$ which will yield a finite contribution to the upper bound. In the second region, $\left|k-i\right|>2v_{\!L\!R}t$, the causal-cones are well separated so we will be able to use the exponential clustering to produce a finite upper bound. The contributions from both regions will be summed up in the end to get the overall upper bound on the OTOC. We will treat $c_{ijk}(t)$ slightly differently in the two regions:
\begin{eqnarray}
c_{ijk}(t)&:=&-\left\langle \left[w_{i}(t),v_{j}\right],\left[w_{k}(t),v_{0}\right]\right\rangle _{\beta}^{c}\nonumber\\
&=&\begin{cases}
-\left\langle \left[w_{i}(t),v_{j}\right],\left[w_{k}(t),v_{0}\right]\right\rangle _{\beta}^{c}; & \left|k-i\right|\leq 2v_{\!L\!R}t\\
-\left\langle \left[\left.w_{i}(t)\right|_{\Gamma_{i}}+w_{i}(t)-\left.w_{i}(t)\right|_{\Gamma_{i}},v_{j}\right],\left[\left.w_{k}(t)\right|_{\Gamma_{k}}+w_{k}(t)-\left.w_{k}(t)\right|_{\Gamma_{k}},v_{0}\right]\right\rangle _{\beta}^{c}; & \left|k-i\right|>2v_{\!L\!R}t
\end{cases}.
\end{eqnarray}
In the proof, the following obvious bound on the connected correlation functions will be useful,
\be
|\ave{a,b}_\beta^{c}|\leq|\ave{ab}_\beta|+|\ave{a}_\beta\ave{b}_\beta| \leq 2\left\|a\right\|\left\|b\right\|.\label{eq:NormConnectedApp}
\ee
We will also need a rigorous estimate for the norm of a projected operator $\left.a\right|_{\Gamma}=a-\left(a-\left.a\right|_{\Gamma}\right)$:
\begin{eqnarray}
\left\|\left.a_x(t)\right|_{\Gamma}\right\|&\leq&\left\|a_x(t)\right\|+\left\|a_x(t)-\left.a_x(t)\right|_{\Gamma}\right\|\nonumber\\
&\leq&\left\|a_x\right\|+\xi\left\|a_x\right\|e^{-\mu\max\left\{0,\dd\left(x,\Gamma^C\right)-v_{\!L\!R}t\right\}}\nonumber\\
&\leq&\left(1+\xi\right)\left\|a\right\|.\label{eq:NormProjectedApp}
\end{eqnarray}
In the first line, we used the triangular inequality. In the second line, we utilised the fact that unitary time evolution preserves the norm, followed by an application of the second version of the LRT, Eq.~\eqref{eq:LiebRobinson2App}. In the last line, we used the fact that the exponential of a non-positive function can be bound by 1.

\noindent\underline{\textbf{For $\left|k-i\right|\leq2v_{\!L\!R}t$ :}}
\\
\begin{eqnarray}
\left|c_{ijk}(t)\right|	&\leq&	2\left\Vert \left[w_{i}(t),v_{j}\right]\right\Vert \left\Vert \left[w_{k}(t),v_{0}\right]\right\Vert \nonumber\\
&\leq&	2\,\xi^{2}\left\Vert w\right\Vert ^{2}\left\Vert v\right\Vert ^{2}e^{-\mu\max\left(0,\left|i-j\right|-v_{\!L\!R}t\right)}e^{-\mu\max\left(0,\left|k\right|-v_{\!L\!R}t\right)},
\end{eqnarray}
where we used Eq. \eqref{eq:NormConnectedApp} to bound the connected correlator and the LRT \eqref{eq:LiebRobinson1App} to bound the norms of the commutators. We can now sum over the corresponding domain to get the contribution to the overall upper bound: 
\begin{eqnarray}
\sum_{k}\hspace{1em}\sum_{i=k-2v_{\!L\!R}t}^{k+2v_{\!L\!R}t}\sum_{j}\left|c_{ijk}(t)\right|	&\leq&	2\,\xi^{2}\left\Vert w\right\Vert ^{2}\left\Vert v\right\Vert ^{2}\sum_{k}\sum_{i=k-2v_{\!L\!R}t}^{k+2v_{\!L\!R}t}\sum_{j}e^{-\mu\max\left(0,\left|i-j\right|-v_{\!L\!R}t\right)}e^{-\mu\max\left(0,\left|k\right|-v_{\!L\!R}t\right)}\nonumber\\
&=&	2\,\xi^{2}\left\Vert w\right\Vert ^{2}\left\Vert v\right\Vert ^{2}\left(\sum_{j}e^{-\mu\max\left(0,\left|i-j\right|-v_{\!L\!R}t\right)}\right)\left(\sum_{k}e^{-\mu\max\left(0,\left|k\right|-v_{\!L\!R}t\right)}\right)\left(\sum_{i=k-2v_{\!L\!R}t}^{k+2v_{\!L\!R}t}1\right)\nonumber\\
&=&	2\,\xi^{2}\left\Vert w\right\Vert ^{2}\left\Vert v\right\Vert ^{2}\left(2v_{\!L\!R}t+\coth\left(\frac{\mu}{2}\right)\right)\left(2v_{\!L\!R}t+\coth\left(\frac{\mu}{2}\right)\right)\left(1+4v_{\!L\!R}t\right)\nonumber\\
&=&	32\,\xi^{2}\left\Vert w\right\Vert ^{2}\left\Vert v\right\Vert ^{2}\left(v_{\!L\!R}t\right)^{3}+\mathcal{O}\left(t^{2}\right).
\end{eqnarray}
In the second line, we used the fact that each individual sum is independent of all other coefficients.
\\

\noindent\underline{\textbf{For $\left|k-i\right|>2v_{\!L\!R}t$ :}}\\
\\
We can now use the linearity of commutators and thermal expectation values to write $c_{ijk}(t)$
as a sum of four terms and then bound it from above, again using the triangular inequality:
\begin{align}
\left|c_{ijk}(t)\right|	\leq& 	\left|\left\langle \left[\left.w_{i}(t)\right|_{\Gamma_{i}},v_{j}\right],\left[\left.w_{k}(t)\right|_{\Gamma_{k}},v_{0}\right]\right\rangle _{\beta}^{c}\right|   && \left(I\right)\nonumber\\
&+\left|\left\langle \left[w_{i}(t)-\left.w_{i}(t)\right|_{\Gamma_{i}},v_{j}\right],\left[\left.w_{k}(t)\right|_{\Gamma_{k}},v_{0}\right]\right\rangle _{\beta}^{c}\right|  && \left(II\right)\nonumber\\
&+\left|\left\langle \left[\left.w_{i}(t)\right|_{\Gamma_{i}},v_{j}\right],\left[w_{k}(t)-\left.w_{k}(t)\right|_{\Gamma_{k}},v_{0}\right]\right\rangle _{\beta}^{c}\right|  &&\left(III\right)\nonumber\\
&+\left|\left\langle \left[w_{i}(t)-\left.w_{i}(t)\right|_{\Gamma_{i}},v_{j}\right],\left[w_{k}(t)-\left.w_{k}(t)\right|_{\Gamma_{k}},v_{0}\right]\right\rangle _{\beta}^{c}\right|. &&\left(IV\right)
\end{align}
Let us now find the upper bounds for each of the terms individually.

\underline{Term $\left(I\right)$:}
Since the supports of the two commutators are well separated (the first commutator is different from identity only on $\Gamma_{i}$, the second one only on $\Gamma_{k}$), we can use the exponential clustering property of thermal states to bound the term. This is the only place in the proof where this property of thermal states is used. However, here, it is indeed crucial:
\begin{eqnarray}
&&\left|\left\langle \left[\left.w_{i}(t)\right|_{\Gamma_{i}},v_{j}\right],\left[\left.w_{k}(t)\right|_{\Gamma_{k}},v_{0}\right]\right\rangle _{\beta}^c\right|\nonumber\\
&&\hspace{1cm}\leq	\chi\left\Vert \left[\left.w_{i}(t)\right|_{\Gamma_{i}},v_{j}\right]\right\Vert \left\Vert \left[\left.w_{k}(t)\right|_{\Gamma_{k}},v_{0}\right]\right\Vert e^{-\rho\,\text{d}(\Gamma_{k},\Gamma_{i})}\nonumber\\
&&\hspace{1cm}\leq	\chi\xi^{2}\left(1+\xi\right)^2\left\Vert w\right\Vert ^{2}\left\Vert v\right\Vert ^{2}e^{-\mu\max\left(0,\left|i-j\right|-v_{\!L\!R}t\right)}\Theta\left(j\in\Gamma_{i}\right)e^{-\mu\max\left(0,\left|k\right|-v_{\!L\!R}t\right)}\Theta\left(0\in\Gamma_{k}\right)e^{-\rho\,\dd(\Gamma_{i},\Gamma_{k})}\nonumber\\
&&\hspace{1cm}\leq	\chi\xi^{2}\left(1+\xi\right)^2\left\Vert w\right\Vert ^{2}\left\Vert v\right\Vert ^{2}e^{-\mu\max\left(0,\left|i-j\right|-v_{\!L\!R}t\right)}e^{-\mu\max\left(0,\left|k\right|-v_{\!L\!R}t\right)}e^{-\rho\,\dd(\Gamma_{i},\Gamma_{k})}\nonumber\\
&&\hspace{1cm}=:	\text{bound[I]}_{ijk}.\label{eq:TermIapp}
\end{eqnarray}
In the first line, we used the exponential clustering property. In the second line, we used the LRT together with \eqref{eq:NormProjectedApp} and the fact that the commutator is non-zero only if $v_0$ and $v_j$ are located inside the supports of $w_i(t)|_{\Gamma_i}$ and $w_k(t)|_{\Gamma_k}$, respectively. In the third line, we used $\Theta\left(\bullet\right)\leq1$, where $\Theta$ is defined as $\Theta(\text{true}) = 1$, $\Theta(\text{false}) = 0$.

Note that if one wanted to compute the density of the disconnected OTOC, this bound would still be valid, but only in the infinite temperature regime $\beta=0$, where the expectation values of the commutators vanish because of the cyclicity of the trace. At finite temperature, an estimate obtained using exponential clustering gives a divergent contribution upon summation over $\sum_{\left|k-i\right|>2v_{\!L\!R}t}\sum_{j}$, indicating that the disconnected dOTOC is generically not a well defined quantity in the thermodynamic limit.

\underline{Term $\left(II\right)$:}
Here, we will use \eqref{eq:NormConnectedApp} to bound the term. To obtain the bound, which will give a non-divergent contribution when summed over $\sum_{\left|k-i\right|>2v_{\!L\!R}t}\sum_{j}$, we will take advantage of a convenient fact that the first commutator can be bound in two different ways---using two different versions of the LRT. One version will give us exponential damping when $\left|k-i\right|$ grows to infinity, the other version when $\left|i-j\right|$ grows to infinity. For a given combination of $i,j,k$, we then take the minimum of the two bounds, which results in a convergent bound upon summation over the domain:
\begin{eqnarray}
&&\left|\left\langle \left[w_{i}(t)-\left.w_{i}(t)\right|_{\Gamma_{i}},v_{j}\right],\left[\left.w_{k}(t)\right|_{\Gamma_{k}},v_{0}\right]\right\rangle _{\beta}^c\right|\nonumber	\\
&&\hspace{1cm}\leq	2\min\left\{ \begin{array}{c}
\left\Vert w_{i}(t)-\left.w_{i}(t)\right|_{\Gamma_{i}}\right\Vert \left\Vert v_{j}\right\Vert \\
\left\Vert \left[w_{i}(t),v_{j}\right]\right\Vert +\left\Vert \left[\left.w_{i}(t)\right|_{\Gamma_{i}},v_{j}\right]\right\Vert 
\end{array}\right\} \left\Vert \left[\left.w_{k}(t)\right|_{\Gamma_{k}},v_{0}\right]\right\Vert\nonumber\\ 
&&\hspace{1cm}\leq	2\,\xi^{2}\left(1+\xi\right)\left\Vert w\right\Vert ^{2}\left\Vert v\right\Vert ^{2}\min\left\{ \begin{array}{c}
e^{-\mu\left(\dd(i,\Gamma_{i}^C)-v_{\!L\!R}t\right)}\\
\left(1+\left(1+\xi\right)\Theta\left(j\in\Gamma_{i}\right)\right)e^{-\mu\max\left(0,\left|i-j\right|-v_{\!L\!R}t\right)}
\end{array}\right\} e^{-\mu\max\left(0,\left|k\right|-v_{\!L\!R}t\right)}\Theta\left(0\in\Gamma_{k}\right)\nonumber\\
&&\hspace{1cm}\leq	2\,\xi^{2}\left(2+\xi\right)\left(1+\xi\right)\left\Vert w\right\Vert ^{2}\left\Vert v\right\Vert ^{2}\min\left\{ \begin{array}{c}
e^{-\mu\left(\dd(i,\Gamma_{i}^C)-v_{\!L\!R}t\right)}\\
e^{-\mu\max\left(0,\left|i-j\right|-v_{\!L\!R}t\right)}
\end{array}\right\} e^{-\mu\max\left(0,\left|k\right|-v_{\!L\!R}t\right)}\nonumber\\
&&\hspace{1cm}=	2\,\xi^{2}\left(2+\xi\right)\left(1+\xi\right)\left\Vert w\right\Vert ^{2}\left\Vert v\right\Vert ^{2}\exp\left[-\mu\max\left\{ \begin{array}{c}
\dd(i,\Gamma_{i}^C)-v_{\!L\!R}t\\
\max\left(0,\left|i-j\right|-v_{\!L\!R}t\right)
\end{array}\right\} \right]e^{-\mu\max\left(0,\left|k\right|-v_{\!L\!R}t\right)}\nonumber\\
&&\hspace{1cm}=:	\text{bound[II]}_{ijk}  . \label{eq:TermIIapp}
\end{eqnarray}
Since we only want to prove that the term will contribute to the upper bound no more (no faster) than polynomially in time, we were allowed to make some of the terms in the third line larger by a constant factor. In the fourth line, we used the fact that the functions appearing in the exponent next to $-\mu$ are non-negative.\\

\underline{Term $\left(III\right)$:}
In analogy with the previous term: 
\begin{eqnarray}
&&\left|\left\langle \left[\left.w_{i}(t)\right|_{\Gamma_{i}},v_{j}\right],\left[w_{k}(t)-\left.w_{k}(t)\right|_{\Gamma_{k}},v_{0}\right]\right\rangle _{\beta}^c\right|\nonumber\\
&&\hspace{1cm}\leq	2\left\Vert \left[\left.w_{i}(t)\right|_{\Gamma_{i}},v_{j}\right]\right\Vert \min\left\{ \begin{array}{c}
\left\Vert w_{k}(t)-\left.w_{k}(t)\right|_{\Gamma_{k}}\right\Vert \left\Vert v_{0}\right\Vert \\
\left\Vert \left[w_{k}(t),v_{0}\right]\right\Vert +\left\Vert \left[\left.w_{k}(t)\right|_{\Gamma_{k}},v_{0}\right]\right\Vert 
\end{array}\right\}\nonumber\\
&&\hspace{1cm}\leq	2\,\xi^{2}\left(1+\xi\right)\left\Vert w\right\Vert ^{2}\left\Vert v\right\Vert ^{2}e^{-\mu\max\left(0,\left|i-j\right|-v_{\!L\!R}t\right)}\Theta\left(j\in\Gamma_{i}\right)\min\left\{ \begin{array}{c}
e^{-\mu\left(\dd(k,\Gamma_{k}^C)-v_{\!L\!R}t\right)}\\
\left(1+\left(1+\xi\right)\Theta\left(0\in\Gamma_{k}\right)\right)e^{-\mu\max\left(0,\left|k\right|-v_{\!L\!R}t\right)}
\end{array}\right\}\nonumber\\
&&\hspace{1cm}\leq	2\,\xi^{2}\left(1+\xi\right)\left(2+\xi\right)\left\Vert w\right\Vert ^{2}\left\Vert v\right\Vert ^{2}e^{-\mu\max\left(0,\left|i-j\right|-v_{\!L\!R}t\right)}\exp\left[-\mu\max\left\{ \begin{array}{c}
\dd(k,\Gamma_{k}^C)-v_{\!L\!R}t\\
\max\left(0,\left|k\right|-v_{\!L\!R}t\right)
\end{array}\right\} \right]\nonumber\\
&&\hspace{1cm}=:	\text{bound[III]}_{ijk}.\label{eq:TermIIIapp}
\end{eqnarray}\\

\underline{Term $\left(IV\right)$:}
Writing again all possible combinations of different versions of the LRT and taking the minimum: 
\begin{eqnarray}
&&\left|\left\langle \left[w_{i}(t)-\left.w_{i}(t)\right|_{\Gamma_{i}},v_{j}\right],\left[w_{k}(t)-\left.w_{k}(t)\right|_{\Gamma_{k}},v_{0}\right]\right\rangle _{\beta}^c\right|\nonumber\\
&&\hspace{1cm}\leq	2\min\left\{ \begin{array}{c}
\left\Vert w_{i}(t)-\left.w_{i}(t)\right|_{\Gamma_{i}}\right\Vert \left\Vert v_{j}\right\Vert \left\Vert w_{k}(t)-\left.w_{k}(t)\right|_{\Gamma_{k}}\right\Vert \left\Vert v_{0}\right\Vert \\
\left\Vert w_{i}(t)-\left.w_{i}(t)\right|_{\Gamma_{i}}\right\Vert \left\Vert v_{j}\right\Vert \left(\left\Vert \left[w_{k}(t),v_{0}\right]\right\Vert +\left\Vert \left[\left.w_{k}(t)\right|_{\Gamma_{k}},v_{0}\right]\right\Vert \right)\\
\left(\left\Vert \left[w_{i}(t),v_{j}\right]\right\Vert +\left\Vert \left[\left.w_{i}(t)\right|_{\Gamma_{i}},v_{j}\right]\right\Vert \right)\left\Vert w_{k}(t)-\left.w_{k}(t)\right|_{\Gamma_{k}}\right\Vert \left\Vert v_{0}\right\Vert \\
\left(\left\Vert \left[w_{i}(t),v_{j}\right]\right\Vert +\left\Vert \left[\left.w_{i}(t)\right|_{\Gamma_{i}},v_{j}\right]\right\Vert \right)\left(\left\Vert \left[w_{k}(t),v_{0}\right]\right\Vert +\left\Vert \left[\left.w_{k}(t)\right|_{\Gamma_{k}},v_{0}\right]\right\Vert \right)
\end{array}\right\}\nonumber\\
&&\hspace{1cm}\leq	2\,\xi^{2}\left\Vert w\right\Vert ^{2}\left\Vert v\right\Vert ^{2}\min\left\{ \begin{array}{c}
e^{-\mu\left(\dd(i,\Gamma_{i}^C)-v_{\!L\!R}t\right)}e^{-\mu\left(\dd(k,\Gamma_{k}^C)-v_{\!L\!R}t\right)}\\
e^{-\mu\left(\dd(i,\Gamma_{i}^C)-v_{\!L\!R}t\right)}\left(1+\left(1+\xi\right)\Theta\left(0\in\Gamma_{k}\right)\right)e^{-\mu\max\left(0,\left|k\right|-v_{\!L\!R}t\right)}\\
\left(1+\left(1+\xi\right)\Theta\left(j\in\Gamma_{i}\right)\right)e^{-\mu\max\left(0,\left|i-j\right|-v_{\!L\!R}t\right)}e^{-\mu\left(\dd(k,\Gamma_{k}^C)-v_{\!L\!R}t\right)}\\
\left(1+\left(1+\xi\right)\Theta\left(j\in\Gamma_{i}\right)\right)e^{-\mu\max\left(0,\left|i-j\right|-v_{\!L\!R}t\right)}\left(1+\left(1+\xi\right)\Theta\left(0\in\Gamma_{k}\right)\right)e^{-\mu\max\left(0,\left|k\right|-v_{\!L\!R}t\right)}
\end{array}\right\}\nonumber\\
&&\hspace{1cm}\leq	2\,\xi^{2}\left\Vert w\right\Vert ^{2}\left\Vert v\right\Vert ^{2}\min\left\{ \begin{array}{c}
e^{-\mu\left(\dd(i,\Gamma_{i}^C)-v_{\!L\!R}t\right)}e^{-\mu\left(\dd(k,\Gamma_{k}^C)-v_{\!L\!R}t\right)}\\
\left(2+\xi\right)e^{-\mu\left(\dd(i,\Gamma_{i}^C)-v_{\!L\!R}t\right)}e^{-\mu\max\left(0,\left|k\right|-v_{\!L\!R}t\right)}\\
\left(2+\xi\right)e^{-\mu\max\left(0,\left|i-j\right|-v_{\!L\!R}t\right)}e^{-\mu\left(\dd(k,\Gamma_{k}^C)-v_{\!L\!R}t\right)}\\
\left(2+\xi\right)^2e^{-\mu\max\left(0,\left|i-j\right|-v_{\!L\!R}t\right)}e^{-\mu\max\left(0,\left|k\right|-v_{\!L\!R}t\right)}
\end{array}\right\}\nonumber\\
&&\hspace{1cm}\leq	2\,\xi^{2}\left(2+\xi\right)^2\left\Vert w\right\Vert ^{2}\left\Vert v\right\Vert ^{2}e^{-\mu\max\left(0,\left|i-j\right|-v_{\!L\!R}t\right)}\min\left\{ \begin{array}{c}
e^{-\mu\left(\dd(k,\Gamma_{k}^C)-v_{\!L\!R}t\right)}\\
e^{-\mu\max\left(0,\left|k\right|-v_{\!L\!R}t\right)}
\end{array}\right\}\nonumber \\
&&\hspace{1cm}=	2\,\xi^{2}\left(2+\xi\right)^2\left\Vert w\right\Vert ^{2}\left\Vert v\right\Vert ^{2}e^{-\mu\max\left(0,\left|i-j\right|-v_{\!L\!R}t\right)}\exp\left[-\mu\max\left\{ \begin{array}{c}
\dd(k,\Gamma_{k}^C)-v_{\!L\!R}t\\
\max\left(0,\left|k\right|-v_{\!L\!R}t\right)
\end{array}\right\} \right]\nonumber\\
&&\hspace{1cm} =	\frac{\left(2+\xi\right)}{\left(1+\xi\right)}\,\text{bound[III]}_{ijk} . \label{eq:TermIVapp}
\end{eqnarray}
In the fourth line, we used the fact that $\min$ cannot decrease if we simply omit a couple of (non-negative) functions and if we multiply some of the remaining functions by a constant factor. We have taken the common term of the two remaining functions out of the minimum.

We now have the estimates for all of the four terms so we are ready to sum them over the domain to get the contribution to the bound for the dOTOC. Since the setting is reflection symmetric (upon exchanging $i$ and $k$), it is enough to compute (all the terms) for $k>i+2v_{\!L\!R}t$ and double the result. In the case of $k>i+2v_{\!L\!R}t$, the subsets are:
\begin{eqnarray}
\Gamma_{i}&=&\left(-\infty,i+v_{\!L\!R}t+\frac{k-i-2v_{\!L\!R}t}{3}\right]=\left(-\infty,\frac{2}{3}i+\frac{1}{3}k+\frac{1}{3}v_{\!L\!R}t\right],\nonumber\\
\Gamma_{k}&=&\left[k-v_{\!L\!R}t-\frac{k-i-2v_{\!L\!R}t}{3},\infty\right)=\left[\frac{2}{3}k+\frac{1}{3}i-\frac{1}{3}v_{\!L\!R}t,\infty\right),
\end{eqnarray}
and the distances:
\begin{eqnarray}
\dd(i,\Gamma_{i}^C) &=&\dd(k,\Gamma_{k}^C)=\frac{1}{3}\left(k-i+v_{\!L\!R}t\right),\nonumber\\
\dd(\Gamma_{i},\Gamma_{k})&=&\frac{1}{3}\left(k-i-2v_{\!L\!R}t\right).
\end{eqnarray}
We can plug these into the expressions (\ref{eq:TermIapp}-\ref{eq:TermIVapp}) and evaluate the sums (all in the form of geometric series). We find that the contribution to the upper bound on the dOTOC coming from the region $\left|k-i\right|>2v_{\!L\!R}t$ is of order of $t^2$:
\be
2\sum_{k}\,\sum_{i<k-2v_{\!L\!R}t}\,\sum_{j}\left(\text{bound[I]}_{ijk}+\text{bound[II]}_{ijk}+\left(1+\frac{\left(2+\xi\right)}{\left(1+\xi\right)}\right)\text{bound[III]}_{ijk}\right)=\mathcal{O}\left(t^{2}\right),
\ee
since each of the bounds gives an $\mathcal{O}\left(t^{2}\right)$ contribution upon the summation.

By adding this that to the result for $\left|k-i\right|\leq2v_{\!L\!R}t$, we arrive to the end of the proof. Hence, we have established that the dOTOC cannot grow faster than with the third power of time:
\begin{eqnarray}
c(t)&\leq& \sum_{\left|k-i\right|\leq2v_{\!L\!R}t}\sum_j \left|c_{ijk}(t)\right|+\sum_{\left|k-i\right|>2v_{\!L\!R}t}\sum_j \left|c_{ijk}(t)\right|\nonumber\\
&\leq&32\,\xi^{2}\left\Vert w\right\Vert ^{2}\left\Vert v\right\Vert ^{2}v_{\!L\!R}^{3}\,t^{3}+\mathcal{O}\left(t^{2}\right).
\end{eqnarray}

This result can be straightforwardly extended in two ways:
\begin{itemize}
\item
Even without taking the thermodynamic limit in (\ref{eq:ExtensiveOTOCapp}), we can still find a bound on $c^{(N)}(t)$ by using exactly the same formal steps. We only have to assume that the finite $N$ lattice is periodic, so translational invariance can be used.

\item We may consider any regular $d$-dimensional lattice in the regime where the temperature is sufficiently high for the generalisation of the exponential clustering property to hold \cite{Eisert}. In this case, each summation over a positional index (with an appropriate constraint) yields a factor that scales as ${\cal O}(t^d)$, rather than ${\cal O}(t)$. With these results in hand, we finally arrive at the general polynomial bound stated in Eq. \eqref{eq:polbound} of the main text.
\end{itemize}

\section{Numerical evaluation of OTOC based on typicality}
\label{app:Typicality}

The approximative numerical method for evaluating the OTOC for intermediate system sizes, $N\sim22$, that we used is based on Levy's lemma (also referred to as the measure concentration, or typicality). The lemma states, roughly, that in a large enough Hilbert space, the expectation value of a well-behaved observable on a single randomly chosen quantum state will be exponentially close in probability to the ensemble average of the observable. That is, in a large enough Hilbert space, almost any state is typical. For a precise formulation, see for example Refs.~\cite{GerkenLevyLema2013,Mueller2011}. Here, we will approximate ensemble averages by averaging over a set $\left\lbrace\left|\Psi_{\text{rand}}\right\rangle\right\rbrace$ of random states in the Hilbert space. In this case, for an observable {A}, typicality arguments lead to
\be
\left\langle a\right\rangle_{\beta=0}\approx\frac{1}{\left|\left\lbrace\left|\Psi_{\text{rand}}\right\rangle\right\rbrace\right|}\sum_{\left\lbrace\left|\Psi_{\text{rand}}\right\rangle\right\rbrace}\left\langle\Psi_{\text{rand}}\right| a \left|\Psi_{\text{rand}}\right\rangle,
\ee
where $|{\cal S}|$ denotes the cardinality of the set ${\cal S}$ (i.e. the number of random states used in the calculation). Rather than estimating the error of such an approximation by analytical arguments, we will estimate it numerically by computing variances.  

The numerical method for computing the OTOC is then constructed as follows. We generate two sets $\left\lbrace\left|\Psi_1\right\rangle\right\rbrace$ and $\left\lbrace\left|\Psi_2\right\rangle\right\rbrace$ of random (normalised) vectors in the $2^N$ dimensional Hilbert space. Then, we can compute
\begin{align}
\left\langle W(t)VW(t)V\right\rangle_{\beta=0}  &\approx     \frac{1}{\left|\left\lbrace\left|\Psi_1\right\rangle\right\rbrace\right|}\sum_{\left\lbrace\left|\Psi_1\right\rangle\right\rbrace}\left\langle\Psi_1\right| W(t)VW(t)V \left|\Psi_1\right\rangle\nonumber\\
&\approx\frac{1}{\left|\left\lbrace\left|\Psi_1\right\rangle\right\rbrace\right|}\frac{2^N}{\left|\left\lbrace\left|\Psi_2\right\rangle\right\rbrace\right|}\sum_{\left\lbrace\left|\Psi_1\right\rangle\right\rbrace}\sum_{\left\lbrace\left|\Psi_2\right\rangle\right\rbrace}\left\langle\Psi_1\right| W(t)V\left|\Psi_2\right\rangle \left\langle\Psi_2\right| W(t) V \left|\Psi_1\right\rangle\nonumber\\
&=\frac{1}{\left|\left\lbrace\left|\Psi_1\right\rangle\right\rbrace\right|}\frac{2^N}{\left|\left\lbrace\left|\Psi_2\right\rangle\right\rbrace\right|}\sum_{\left\lbrace\left|\Psi_1\right\rangle\right\rbrace}\sum_{\left\lbrace\left|\Psi_2\right\rangle\right\rbrace}     \left\langle\Psi_1(t)\right| W \, |\widetilde{\Psi_2}(t) \rangle \left\langle\Psi_2(t)\right| W \, | \widetilde{\Psi_1}(t) \rangle.\label{eq:OTOCapprox}
\end{align}
In the second line, we inserted a partition of unity and approximated it by $\mathds{1}=\sum_{\left|\Psi\right\rangle\in\mathcal{H}}\left|\Psi\right\rangle \left\langle\Psi\right|\approx\frac{2^N}{\left|\left\lbrace\left|\Psi\right\rangle\right\rbrace\right|}\sum_{\left\lbrace\left|\Psi\right\rangle\right\rbrace}\left|\Psi\right\rangle \left\langle\Psi\right|$. In the third line, we defined $|\widetilde{\Psi} \rangle:=V\left|\Psi\right\rangle$ and switched from the Heisenberg to the Schr\"{o}dinger picture. A similar expression can be derived for the other term appearing in the OTOC, namely $\left\langle V W(t)W(t)V\right\rangle_{\beta=0}=\left\langle V^2 W^2(t)\right\rangle_{\beta=0}$.  

In evaluating the dynamics of $|\Psi_{1,2}(t)\rangle$, the computation of the action of one-spin (or two-spin) unitary operators on a vector is numerically very efficient. This is because the operators act only on two (or four) among the composite spin indices of the vector at the time. Thus, the operation only requires $2\cdot2^N$ (or $4\cdot2^N$) computational steps. The evaluation of the expression \eqref{eq:OTOCapprox} is composed only of one-spin and two-spin operations and can therefore be computed within $\mathcal{O}\left(N2^N\right)$ computational steps. For a detailed discussion of such an algorithm, see for example \cite{ProsenJPA2007}.

Finally, the statistical error of such an approximation is estimated by
\be
\Delta c^{(N)}(t)=\frac{\sigma_{c^{(N)}(t)}}{\sqrt{\left|\left\lbrace\left|\Psi_1\right\rangle\right\rbrace\right|\left|\left\lbrace\left|\Psi_2\right\rangle\right\rbrace\right|}},
\ee
where $\sigma_{c^{(N)}}$ is the variance of $\left\lbrace c^{(N)}_{\left|\Psi_1\right\rangle,\left|\Psi_2\right\rangle}\right\rbrace$ if 
$c^{(N)}_{\left|\Psi_1\right\rangle,\left|\Psi_2\right\rangle}$ is an expression like RHS of Eq. \eqref{eq:OTOCapprox} for a fixed state pair $\left|\Psi_1\right\rangle,\left|\Psi_2\right\rangle$ (omitting the averaging).

\section{Kicked quantum Ising propagator in the Fourier basis}\label{app:KIFourier}
\label{apKI}

Propagators of translationally invariant free models can be simplified by writing them in the Fourier transformed basis of Majorana fermions. This is useful for computing the action of powers of the propagator and therefore obtaining real time dynamics.

The Fourier transformed Majorana fermions can be written as
\begin{align}
w(\theta)=\sum_{j}w_{2j}e^{\ii \theta j}, && w'(\theta)=\sum_{j}w_{2j+1}e^{\ii \theta j}\label{eq:Fourier}.
\end{align}
In most of the calculations here, we can safely assume that the chain is infinite, $N=\infty$, and hence, $\theta \in [-\pi,\pi)$ is a continuous quasi-momentum parameter. It is convenient to introduce the shorthand spinor notation
\begin{equation}
\underline{w}(\theta)=\left(
\begin{matrix}
w(\theta)\\
w'(\theta)
\end{matrix}
\right).
\end{equation}
Written in the Heisenberg picture, the propagator in this basis acts as a $2\times 2$ unitary matrix
\begin{equation}
\mathcal{U}\underline{w}(\theta):=\left(
\begin{matrix}
U^\dagger w(\theta) U\\
U^\dagger w'(\theta) U
\end{matrix}
\right). \label{eq:Heisenberg}
\end{equation}

As an example, let us explicitly compute the expression for the propagator of the transverse KI field ($\varphi=0$) in the Fourier basis:
\begin{eqnarray}
U&=&e^{-J\sum_jw_{2j-1}w_{2j}}e^{-h\sum_jw_{2j}w_{2j+1}}\\
&=&\prod_j\left(\cos\left(J\right)-w_{2j-1}w_{2j}\sin\left(J\right)\right)\prod_k\left(\cos\left(h\right)-w_{2k}w_{2k+1}\sin\left(h\right)\right)\\
&=&U_{\text{Ising}}U_{\text{kick}}.\label{eq:FloquetFermionApp}
\end{eqnarray}
The kick term acts as
\begin{eqnarray}
\mathcal{U}_{\text{kick}}\underline{w}(\theta)&=&\left(
\begin{matrix}
U_{\text{kick}}^\dagger w(\theta) U_{\text{kick}}\\
U_{\text{kick}}^\dagger w'(\theta) U_{\text{kick}}
\end{matrix}
\right)\nonumber\\
&=&\sum_j \left(
\begin{matrix}
U_{\text{kick}}^\dagger w_{2j} U_{\text{kick}}\\
U_{\text{kick}}^\dagger w_{2j+1} U_{\text{kick}}
\end{matrix}
\right)e^{\ii \theta j}\nonumber\\
&=&\sum_j \left(
\begin{matrix}
\left(\cos\left(h\right)+w_{2j}w_{2j+1}\sin\left(h\right)\right) w_{2j} \left(\cos\left(h\right)-w_{2j}w_{2j+1}\sin\left(h\right)\right)\\
\left(\cos\left(h\right)+w_{2j}w_{2j+1}\sin\left(h\right)\right) w_{2j+1} \left(\cos\left(h\right)-w_{2j}w_{2j+1}\sin\left(h\right)\right)
\end{matrix}
\right)e^{\ii \theta j}\nonumber\\
&=&\sum_j \left(
\begin{matrix}
\left(\cos^2\left(h\right)-\sin^2\left(h\right)\right)w_{2j}-2\sin\left(h\right)\cos\left(h\right)w_{2j+1}\\
2\sin\left(h\right)\cos\left(h\right)w_{2j}+\left(\cos^2\left(h\right)-\sin^2\left(h\right)\right)w_{2j+1}
\end{matrix}
\right)e^{\ii \theta j}\nonumber
\\
&=&\left(
\begin{matrix}
\cos(2h) & -\sin(2h)\\
\sin(2h) & \cos(2h)
\end{matrix} \right) \underline{w}(\theta),
\end{eqnarray}
where in the first line, we used the definition of the propagator in the Heisenberg picture \eqref{eq:Heisenberg}. In the second line, we employed the definition of the Fourier transformed Majorana fermions \eqref{eq:Fourier}. Then, in the third line, we used the fact that $w_i$ has a non-trivial product only with terms containing $w_i$ (this follows from the Majorana anti-commutation relations, $\left\{w_i,w_j\right\}=2\delta_{ij}$) and finally, in the last line, we again utilised the definition of the Fourier transform.

Similarly, for the Ising propagator,
\begin{eqnarray}
\mathcal{U}_{\text{Ising}}\underline{w}(\theta)&=&\sum_j\left(
\begin{matrix}
U_{\text{Ising}}^\dagger w_{2j} U_{\text{Ising}}\\
U_{\text{Ising}}^\dagger w_{2j+1} U_{\text{Ising}}
\end{matrix}
\right)e^{\ii \theta j}\nonumber\\
&=&\sum_j\left(
\begin{matrix}
U_{\text{Ising}}^\dagger w_{2j} U_{\text{Ising}} \, e^{\ii \theta j}\\
U_{\text{Ising}}^\dagger w_{2j-1} U_{\text{Ising}} \, e^{\ii \theta \left(j-1\right)}
\end{matrix}
\right)\nonumber\\
&=&\sum_j \left(
\begin{matrix}
\left(\cos\left(J\right)+w_{2j-1}w_{2j}\sin\left(J\right)\right) w_{2j} \left(\cos\left(J\right)-w_{2j-1}w_{2j}\sin\left(J\right)\right) \, e^{\ii \theta j}\\
\left(\cos\left(J\right)+w_{2j-1}w_{2j}\sin\left(J\right)\right) w_{2j-1} \left(\cos\left(J\right)-w_{2j-1}w_{2j}\sin\left(J\right)\right)\, e^{\ii \theta \left(j-1\right)}
\end{matrix}
\right)\nonumber\\
&=&\left(
\begin{matrix}
\cos(2J) & e^{\ii \theta}\sin(2J)\\
-e^{-\ii \theta}\sin(2J) & \cos(2J)
\end{matrix} \right) \underline{w}(\theta).
\end{eqnarray}
We note that in the second line, we shifted the summation index. 

Then, for the action of the entire Floquet propagator (note the correct order), we recover Eq. \eqref{eq:KIFloquetFourier} from the main text:
\begin{eqnarray}
\mathcal{U}=\mathcal{U}_{\text{kick}}\mathcal{U}_{\text{Ising}} = \left(
\begin{matrix}
\cos(2h) & -\sin(2h)\\
\sin(2h) & \cos(2h)
\end{matrix} \right)
\left(
\begin{matrix}
\cos(2J) & e^{\ii \theta}\sin(2J)\\
-e^{-\ii \theta}\sin(2J) & \cos(2J)
\end{matrix} \right).\label{eq:KIFloquetFourierApp}
\end{eqnarray}

\section{Spectrum of the kicked quantum Ising model} \label{app:SpectrumKI}

The Floquet propagator $\mathcal{U}$ of a general quadratic model in the Fourier basis is a $2\times2$ unitary matrix. We can diagonalise it to the following form:
\begin{equation}
\mathcal{U}(\theta)=V^\dagger(\theta) \left(\begin{matrix}
e^{\ii \kappa(\theta)} && \\
&& e^{\ii \lambda(\theta)}
\end{matrix}\right) V(\theta),\label{eq:FloquetUnitary}
\end{equation}
where $V$ is a unitary eigenvector matrix with elements
\begin{equation}
V(\theta)=\left(\begin{matrix}
v_{11}(\theta) && v_{12}(\theta) \\
v_{21}(\theta) && v_{22}(\theta)
\end{matrix}\right).\label{eq:EigenV}
\end{equation}

\begin{figure}
	\includegraphics[width=0.5\linewidth]{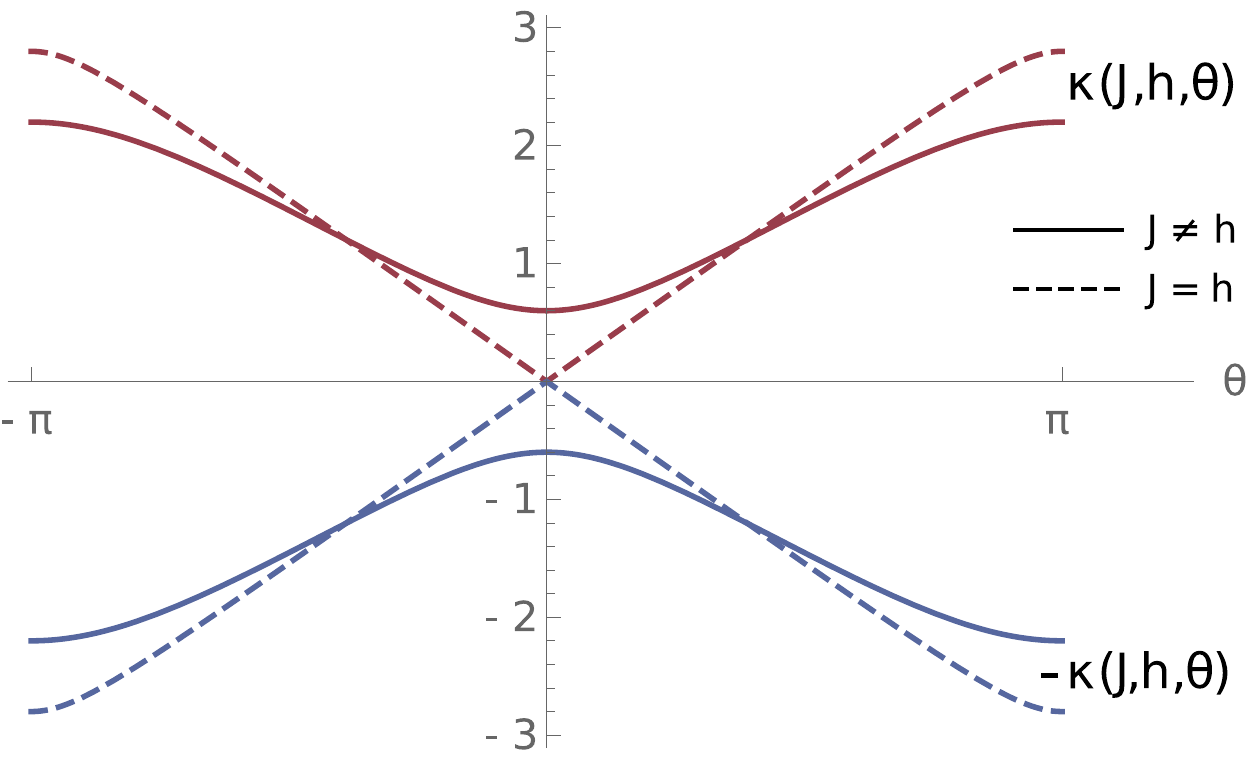}
	\caption{\label{fig:SpectrumKI} Floquet quasiparticle spectrum (eigenphases) of the kicked quantum Ising model \eqref{eq:EigenvalueSupp}, \eqref{eq:MinusEigenvalueSupp}. The full lines represents generic curves for the case of $J\neq h$, for which the spectrum has a gap. The dashed lines represent generic curves for the case of $J=h$, for which the gap closes and the system exhibits a Floquet analogue of a quantum phase transition.}
\end{figure}

Diagonalising \eqref{eq:KIFloquetFourierApp} and using $\arccos\left(z\right)=-\ii\ln\left(z+\sqrt{z^2-1}\right)$,
we find the eigenvalues to be 
\begin{align}
\kappa(J,h,\theta) &=\arccos\left[\cos(2J)\cos(2h)+\cos(\theta)\sin(2J)\sin(2h)\right], \label{eq:EigenvalueSupp} \\
\lambda(J,h,\theta)&=-\kappa(J,h,\theta).\label{eq:MinusEigenvalueSupp}
\end{align}
We see that the Floquet quasiparticle dispersion relation $\kappa(J,h,\theta)$ has three extrema: two maxima at $\theta=\pm\pi$ and a minimum at $\theta=0$. For $J\neq h$, we have $\kappa\left(J\neq h,\theta\right)>0$ so the system has a spectral gap (see Fig. \ref{fig:SpectrumKI}). For $J=h$, $\kappa\left(J=h,\theta=0\right)=-\kappa\left(J=h,\theta=0\right)=0$, and so the gap closes. This is a Floquet-type analogue of a quantum critical line in the $J$--$h$ plane. On the $J=h$ line, the eigenphase $\kappa\left(J=h,\theta\right)$ has only two extrema at $\theta=\pm\pi$.

For the eigenvectors, we find
\begin{align}
v_{11}(J,h,\theta) & =
\frac{e^{-\ii \theta
	}\sin(2J)\sin(2h)+\cos(2J)\cos(2h)-e^{-\ii \kappa(J,h,\theta)}}
{e^{\ii \theta
	}\sin(2J)\cos(2h)-\cos(2J)\sin(2h)}/\text{norm}_1(J,h,\theta), \\
v_{12}(J,h,\theta)&= 1/\text{norm}_1(J,h,\theta), \\
v_{21}(J,h,\theta) &=
\frac{e^{-\ii \theta
	}\sin(2J)\sin(2h)+\cos(2J)\cos(2h)-e^{\ii \kappa(J,h,\theta)}}
{e^{\ii \theta
	}\sin(2J)\cos(2h)-\cos(2J)\sin(2h)}/\text{norm}_2(J,h,\theta), \\
v_{22}(J,h,\theta)&=1/\text{norm}_2(J,h,\theta),
\end{align}
where
\begin{align}
\text{norm}_1(J,h,\theta)&=\sqrt{1+\frac{2\left[\sin\left(\kappa[J,h,\theta]\right)-\sin(\theta)\sin(2J)\sin(2h)\right]^2}{1-\cos\left(\kappa\left[2J,2h,\theta\right]\right)}}, \\
\text{norm}_2(J,h,\theta)&=\sqrt{1+\frac{2\left[\sin\left(\kappa[J,h,\theta]\right)+\sin(\theta)\sin(2J)\sin(2h)\right]^2}{1-\cos\left(\kappa\left[2J,2h,\theta\right]\right)}}.
\end{align}
At the extremal points, $\theta\in\left\{-\pi,0,\pi\right\}$, the eigenvectors simplify to
\begin{equation}
V_{0/\pm\pi}=\left(\begin{matrix}
-\frac{\ii \Sigma_{0/\pm\pi}}{\sqrt{2}} && \frac{1}{\sqrt{2}}\\
\frac{\ii \Sigma_{0/\pm\pi}}{\sqrt{2}} && \frac{1}{\sqrt{2}}
\end{matrix}\right),\label{eq:StationaryVector}
\end{equation}
where
\begin{align}
\Sigma_0=\text{sign}\left[\sin(2h-2J)\right],&&\Sigma_{\pm\pi}=\text{sign}\left[\sin(2h+2J)\right].
\end{align}
Note that $\Sigma_{0/\pm\pi}^2=1$.

\section{Real space propagator}\label{app:Symbol}

The propagator of a general translationally invariant Floquet system, quadratic in fermionic operators (that is, a free model), can be expressed in the Fourier transformed basis as a unitary $2\times2$ matrix that depends on quasi-momentum. The same formalism applies for general time-dependent situation through the application of the Trotter-Suzuki formula. For example, we could obtain the results for the time-independent transverse field Ising model by setting $J\to dt\, J$, $h\to dt \,h$ and then taking the limit  of $dt\to 0$.

By using the expression for $\mathcal{U}(\theta)$ and
\begin{equation}
\underline{w}(\theta,t)=\mathcal{U}(\theta)^t\underline{w}(\theta,0),\label{eq:FourTimeProp}
\end{equation}
we compute the time-evolution of the Majorana fermions in the spatial basis; that is, the real-space propagator $K$, defined by
\begin{equation}
\left(
\begin{matrix}
w_{2j}(t)\\
w_{2j+1}(t)
\end{matrix}
\right)=:\sum_k \left[\left(\begin{matrix}
w_{2k} &
w_{2k+1}
\end{matrix}
\right) K^{kj}(t)\right]^\text{T}
,\label{eq:DefinitionSymbol}
\end{equation}
or equivalently,
\begin{equation}
K_{ab}^{kj}(t):=\left\langle w_{2k+a-1}\,w_{2j+b-1}(t)\right\rangle, \label{eq:NotationSymbolApp}
\end{equation}
for $a,b\in\left\lbrace1,2\right\rbrace$.

Let us start from Eq.~\eqref{eq:FourTimeProp}. Then,
\begin{equation}
\underline{w}(\theta,t)=\mathcal{U}(\theta)^t\underline{w}(\theta).\label{eq:SymDerBegin}
\end{equation}
At the same time, by definition (and by using Eq. \eqref{eq:Fourier}), the above expression is equal to
\begin{equation}
\underline{w}(\theta,t) = \sum_j e^{\ii \theta j}\left(
\begin{matrix}
w_{2j}(t)\\
w_{2j+1}(t)
\end{matrix}
\right)  . \label{eq:DerivSymbolStep1}
\end{equation}
Using the inverse Fourier transform
\begin{equation}
\left(
\begin{matrix}
w_{2k}\\
w_{2k+1}
\end{matrix}
\right)=\frac{1}{2\pi}\int_{-\pi}^{\pi}\dd\varphi\,e^{-\ii \varphi k}\left(
\begin{matrix}
w(\varphi)\\
w'(\varphi)
\end{matrix}
\right),
\end{equation}
along with the definition \eqref{eq:DefinitionSymbol}, the equation \eqref{eq:DerivSymbolStep1} can then be further expressed as
\begin{equation}
=\sum_j e^{\ii \theta j}\sum_k K^{kj}(t)\frac{1}{2\pi}\int_{-\pi}^{\pi}\dd\varphi\,e^{-\ii \varphi k}\left(
\begin{matrix}
w(\varphi)\\
w'(\varphi)
\end{matrix}
\right)=.
\end{equation}
Now, taking into account the translational invariance of the system, i.e. $K^{kj}=:K^{j-k}$, and thereby introducing a new index $l:=j-k$, the expression becomes
\begin{equation}
=\frac{1}{2\pi}\int_{-\pi}^{\pi}\dd\varphi\sum_{k,l}e^{\ii (\theta-\varphi)k}K^l(t) e^{\ii \theta l} \underline{w}(\varphi)=.
\end{equation}
Summing over $k$ and using $\sum_n e^{\ii n x}=2\pi\sum_k\delta\left(x-2\pi k\right)$, we have
\begin{equation}
=\int_{-\pi}^{\pi}\dd\varphi\delta(\theta-\varphi)\sum_l K^l(t) e^{\ii \theta l} \underline{w}(\varphi)=.
\end{equation}
Finally, by integrating over $\varphi$, we get
\begin{equation}
=\sum_l K^l(t) e^{\ii \theta l} \underline{w}(\theta).\label{eq:SymDerEnd}
\end{equation}
Comparing Eqs. \eqref{eq:SymDerBegin} and \eqref{eq:SymDerEnd}, we see that
\begin{equation}
\mathcal{U}(\theta)^t=\sum_l K^l(t) e^{\ii \theta l},
\end{equation}
or equivalently, by performing the inverse Fourier transform
\begin{equation}
K^l(t)=\frac{1}{2\pi}\int_{-\pi}^{\pi}\dd\theta e^{-\ii \theta l} \mathcal{U}^t(\theta).\label{eq:SymbolFinalApp}
\end{equation}

\section{The dOTOC of transverse magnetisation}\label{app:FormulaOTOC}
\label{apOTOC}

We now want to use the general form of the propagator to compute the high temperature limit of the dOTOC for transverse magnetisation in free fermionic systems:
\begin{eqnarray}
c_z(t)&:=&\lim_{N\to\infty}c^{(N)}_z(t)\nonumber\\
&:=&-\frac{1}{N}\left\langle\left[M_z(t),M_z(0)\right]^2\right\rangle_{\beta=0}\nonumber\\
&=&-\frac{2}{N}\left\{\left\langle M_z(t)M_z(0)M_z(t)M_z(0)\right\rangle_{\beta=0}- \left\langle M_z(0)M_z(t)M_z(t)M_z(0)\right\rangle_{\beta=0}\right\}\nonumber\\
&=:&-2\left\{\left(I\right)-\left(II\right)\right\},\label{eq:OTOCmagZ}
\end{eqnarray}
with
\begin{equation}
M_z:=\sum_{j\in\ZZ}\sigma_j^z=-\ii\sum_{j\in\ZZ}w_{2j}w_{2j+1}. \label{eq:MzMajoranaApp}
\end{equation}
Plugging \eqref{eq:MzMajoranaApp} into \eqref{eq:OTOCmagZ} and using the definition \eqref{eq:NotationSymbolApp}, we obtain
\begin{eqnarray}
\left(I\right)&=&\frac{1}{N}\sum_{l_1,j,l_3,k}\left\langle w_{2l_1}(t) w_{2l_1+1}(t)w_{2j}w_{2j+1}w_{2l_3}(t) w_{2l_3+1}(t)w_{2k}w_{2k+1}\right\rangle_{\beta=0}\nonumber\\
&=&\frac{1}{N}\sum_{l_1,j,l_3,k}\sum_{\bar{l}_1,\bar{\bar{l}}_1}\sum_{\bar{s}_1,\bar{\bar{s}}_1=1}^2K_{\bar{s}_1,1}^{l_1-\bar{l}_1}(t)K_{\bar{\bar{s}}_1,2}^{l_1-\bar{\bar{l}}_1}(t)\sum_{\bar{l}_3,\bar{\bar{l}}_3}\sum_{\bar{s}_3,\bar{\bar{s}}_3=1}^2K_{\bar{s}_3,1}^{l_3-\bar{l}_3}(t)K_{\bar{\bar{s}}_3,2}^{l_3-\bar{\bar{l}}_3}(t)\cdot\nonumber\\
&&\cdot\left\langle w_{2\bar{l}_1+\bar{s}_1-1} w_{2\bar{\bar{l}}_1+\bar{\bar{s}}_1-1}w_{2j}w_{2j+1}w_{2\bar{l}_3+\bar{s}_3-1} w_{2\bar{\bar{l}}_3+\bar{\bar{s}}_3-1}w_{2k}w_{2k+1}\right\rangle_{\beta=0}.
\end{eqnarray}
Note that we changed the summation indices compared to those used in the main text ($i$, $k$ to $l_1$, $l_3$ and $l$ to $k$). Taking into account the translational invariance of the KI, we can fix one of the indices in the first sum and replace the summation over that index with an overall multiplication by the number of particles in the system $N$, with $N\rightarrow\infty$. It is convenient to fix the last index to $k=0$.  The expression then simplifies to
\begin{eqnarray}
\left(I\right)&=&\sum_j\sum_{l_1,l_3}\sum_{\bar{l}_1,\bar{\bar{l}}_1}\sum_{\bar{l}_3,\bar{\bar{l}}_3}\sum_{\bar{s}_1,\bar{\bar{s}}_1=1}^2\sum_{\bar{s}_3,\bar{\bar{s}}_3=1}^2K_{\bar{s}_1,1}^{l_1-\bar{l}_1}K_{\bar{\bar{s}}_1,2}^{l_1-\bar{\bar{l}}_1}K_{\bar{s}_3,1}^{l_3-\bar{l}_3}K_{\bar{\bar{s}}_3,2}^{l_3-\bar{\bar{l}}_3}\cdot\nonumber\\
&&\cdot\left\langle w_{2\bar{l}_1+\bar{s}_1-1} w_{2\bar{\bar{l}}_1+\bar{\bar{s}}_1-1}w_{2j}w_{2j+1}w_{2\bar{l}_3+\bar{s}_3-1} w_{2\bar{\bar{l}}_3+\bar{\bar{s}}_3-1}w_{0}w_{1}\right\rangle_{\beta=0},\label{eq:TermI}
\end{eqnarray}
where the temporal dependence of $K$ is omitted for compactness of notation. The expression is formal and will be simplified in what is to follow. Furthermore, we also have
\begin{eqnarray}
\left(II\right)&=&\sum_j\sum_{l_1,l_3}\sum_{\bar{l}_1,\bar{\bar{l}}_1}\sum_{\bar{l}_3,\bar{\bar{l}}_3}\sum_{\bar{s}_1,\bar{\bar{s}}_1=1}^2\sum_{\bar{s}_3,\bar{\bar{s}}_3=1}^2K_{\bar{s}_1,1}^{l_1-\bar{l}_1}K_{\bar{\bar{s}}_1,2}^{l_1-\bar{\bar{l}}_1}K_{\bar{s}_3,1}^{l_3-\bar{l}_3}K_{\bar{\bar{s}}_3,2}^{l_3-\bar{\bar{l}}_3}\cdot\nonumber\\
&&\cdot\left\langle w_{2j}w_{2j+1} w_{2\bar{l}_1+\bar{s}_1-1} w_{2\bar{\bar{l}}_1+\bar{\bar{s}}_1-1}w_{2\bar{l}_3+\bar{s}_3-1} w_{2\bar{\bar{l}}_3+\bar{\bar{s}}_3-1}w_{0}w_{1}\right\rangle_{\beta=0}.\label{eq:TermII}
\end{eqnarray}

The key to simplifying the expressions $\left(I\right)$ and $\left(II\right)$ are the anti-commutation relations 
\be
\left\{w_i,w_j\right\}=2\delta_{ij},\label{eq:MajoranaAntiComm}
\ee
or the equation for the pair correlation function that follows from them:
\begin{equation}
\left\langle w_i \,  w_j\right\rangle_{\beta=0}=\delta_{ij}.\label{eq:TwoPoint}
\end{equation}
We can use it together with the Wick's theorem to compute the eight-fermion correlation functions appearing in \eqref{eq:TermI} and \eqref{eq:TermII}. We see that the infinite temperature expectation values in \eqref{eq:TermI} and \eqref{eq:TermII} are only non-zero if all of the Majorana fermions in them appear in pairs. In particular, it is helpful to consider the cases of $j\neq 0$ and $j=0$ separately. 
\newline

\textbf{a) $j\neq0$} 
\newline

\noindent First, consider the terms in the correlator for $\left(I\right)$ that have the form

\begin{equation}
\left\langle \underline{\,\,\,\,\,\,}\,\, \underline{\,\,\,\,\,\,}w_{2j}w_{2j+1}\underline{\,\,\,\,\,\,}\,\, \underline{\,\,\,\,\,\,}w_{0}w_{1}\right\rangle_{\beta=0},
\end{equation}
where the empty slots ($\underline{\,\,\,\,\,\,}$) have to be filled by $w_{2j}$, $w_{2j+1}$, $w_{0}$, $w_{1}$, each appearing exactly once. This gives $24$ possible permutations. Then, for the corresponding correlator in $\left(II\right)$ (with the first two pairs of terms interchanged), we want
\begin{equation}
\left\langle w_{2j}w_{2j+1} \underline{\,\,\,\,\,\,}\,\, \underline{\,\,\,\,\,\,}\,\,\underline{\,\,\,\,\,\,}\,\, \underline{\,\,\,\,\,\,}w_{0}w_{1}\right\rangle_{\beta=0}=-\left\langle \underline{\,\,\,\,\,\,}\,\, \underline{\,\,\,\,\,\,}w_{2j}w_{2j+1}\underline{\,\,\,\,\,\,}\,\, \underline{\,\,\,\,\,\,}w_{0}w_{1}\right\rangle_{\beta=0},
\end{equation}
so that the combination results in a non-zero term in \eqref{eq:OTOCmagZ}. Note that the only allowed combinations are those where we have only one among $w_{2j}$ and $w_{2j+1}$ in the first two slots. The remaining slot, among the first two slots, has to be filled by either $w_{0}$ or $w_{1}$. This is also a direct consequence of the anti-commutation relations. We are left with $16$ possible permutations, which we write out explicitly:
\begin{eqnarray}
K_{1,1}^{l_1-j}K_{1,2}^{l_1}K_{2,1}^{l_3-j}K_{2,2}^{l_3}
\underbrace{\left\langle w_{2j}w_{0}\,
	w_{2j}w_{2j+1}\,
	w_{2j+1}w_{1}\,
	w_{0}w_{1}\right\rangle_{\beta=0}}_1;
\,\,\,s_j=1,s_0=1,p_1=1,p_3=1,\nonumber
\\
K_{1,1}^{l_1-j}K_{1,2}^{l_1}K_{2,1}^{l_3}K_{2,2}^{l_3-j}
\underbrace{\left\langle w_{2j}w_{0}\,
	w_{2j}w_{2j+1}\,
	w_{1}w_{2j+1}\,
	w_{0}w_{1}\right\rangle_{\beta=0}}_{-1};
\,\,\,s_j=1,s_0=1,p_1=1,p_3=2,\nonumber
\\
K_{1,1}^{l_1}K_{1,2}^{l_1-j}K_{2,1}^{l_3-j}K_{2,2}^{l_3}
\underbrace{\left\langle w_{0}w_{2j}\,
	w_{2j}w_{2j+1}\,
	w_{2j+1}w_{1}\,
	w_{0}w_{1}\right\rangle_{\beta=0}}_{-1};
\,\,\,s_j=1,s_0=1,p_1=2,p_3=1,\nonumber
\\
K_{1,1}^{l_1}K_{1,2}^{l_1-j}K_{2,1}^{l_3}K_{2,2}^{l_3-j}
\underbrace{\left\langle w_{0}w_{2j}\,
	w_{2j}w_{2j+1}\,
	w_{1}w_{2j+1}\,
	w_{0}w_{1}\right\rangle_{\beta=0}}_{1};
\,\,\,s_j=1,s_0=1,p_1=2,p_3=2,\nonumber
\end{eqnarray}

\begin{eqnarray}
K_{1,1}^{l_1-j}K_{2,2}^{l_1}K_{2,1}^{l_3-j}K_{1,2}^{l_3}
\underbrace{\left\langle w_{2j}w_{1}\,
	w_{2j}w_{2j+1}\,
	w_{2j+1}w_{0}\,
	w_{0}w_{1}\right\rangle_{\beta=0}}_{-1};
\,\,\,s_j=1,s_0=2,p_1=1,p_3=1,\nonumber
\\
K_{1,1}^{l_1-j}K_{2,2}^{l_1}K_{1,1}^{l_3}K_{2,2}^{l_3-j}
\underbrace{\left\langle w_{2j}w_{1}\,
	w_{2j}w_{2j+1}\,
	w_{0}w_{2j+1}\,
	w_{0}w_{1}\right\rangle_{\beta=0}}_{1};
\,\,\,s_j=1,s_0=2,p_1=1,p_3=2,\nonumber
\\
K_{2,1}^{l_1}K_{1,2}^{l_1-j}K_{2,1}^{l_3-j}K_{1,2}^{l_3}
\underbrace{\left\langle w_{1}w_{2j}\,
	w_{2j}w_{2j+1}\,
	w_{2j+1}w_{0}\,
	w_{0}w_{1}\right\rangle_{\beta=0}}_{1};
\,\,\,s_j=1,s_0=2,p_1=2,p_3=1,\nonumber
\\
K_{2,1}^{l_1}K_{1,2}^{l_1-j}K_{1,1}^{l_3}K_{2,2}^{l_3-j}
\underbrace{\left\langle w_{1}w_{2j}\,
	w_{2j}w_{2j+1}\,
	w_{0}w_{2j+1}\,
	w_{0}w_{1}\right\rangle_{\beta=0}}_{-1};
\,\,\,s_j=1,s_0=2,p_1=2,p_3=2,\nonumber
\end{eqnarray}

\begin{eqnarray}
K_{2,1}^{l_1-j}K_{1,2}^{l_1}K_{1,1}^{l_3-j}K_{2,2}^{l_3}
\underbrace{\left\langle w_{2j+1}w_{0}\,
	w_{2j}w_{2j+1}\,
	w_{2j}w_{1}\,
	w_{0}w_{1}\right\rangle_{\beta=0}}_{-1};
\,\,\,s_j=2,s_0=1,p_1=1,p_3=1,\nonumber
\\
K_{2,1}^{l_1-j}K_{1,2}^{l_1}K_{2,1}^{l_3}K_{1,2}^{l_3-j}
\underbrace{\left\langle w_{2j+1}w_{0}\,
	w_{2j}w_{2j+1}\,
	w_{1}w_{2j}\,
	w_{0}w_{1}\right\rangle_{\beta=0}}_{1};
\,\,\,s_j=2,s_0=1,p_1=1,p_3=2,\nonumber
\\
K_{1,1}^{l_1}K_{2,2}^{l_1-j}K_{1,1}^{l_3-j}K_{2,2}^{l_3}
\underbrace{\left\langle w_{0}w_{2j+1}\,
	w_{2j}w_{2j+1}\,
	w_{2j}w_{1}\,
	w_{0}w_{1}\right\rangle_{\beta=0}}_{1};
\,\,\,s_j=2,s_0=1,p_1=2,p_3=1,\nonumber
\\
K_{1,1}^{l_1}K_{2,2}^{l_1-j}K_{2,1}^{l_3}K_{1,2}^{l_3-j}
\underbrace{\left\langle w_{0}w_{2j+1}\,
	w_{2j}w_{2j+1}\,
	w_{1}w_{2j}\,
	w_{0}w_{1}\right\rangle_{\beta=0}}_{-1};
\,\,\,s_j=2,s_0=1,p_1=2,p_3=2,\nonumber
\end{eqnarray}

\begin{eqnarray}
K_{2,1}^{l_1-j}K_{2,2}^{l_1}K_{1,1}^{l_3-j}K_{1,2}^{l_3}
\underbrace{\left\langle w_{2j+1}w_{1}\,
	w_{2j}w_{2j+1}\,
	w_{2j}w_{0}\,
	w_{0}w_{1}\right\rangle_{\beta=0}}_{1};
\,\,\,s_j=2,s_0=2,p_1=1,p_3=1,\nonumber
\\
K_{2,1}^{l_1-j}K_{2,2}^{l_1}K_{1,1}^{l_3}K_{1,2}^{l_3-j}
\underbrace{\left\langle w_{2j+1}w_{1}\,
	w_{2j}w_{2j+1}\,
	w_{0}w_{2j}\,
	w_{0}w_{1}\right\rangle_{\beta=0}}_{-1};
\,\,\,s_j=2,s_0=2,p_1=1,p_3=2,\nonumber
\\
K_{2,1}^{l_1}K_{2,2}^{l_1-j}K_{1,1}^{l_3-j}K_{1,2}^{l_3}
\underbrace{\left\langle w_{1}w_{2j+1}\,
	w_{2j}w_{2j+1}\,
	w_{2j}w_{0}\,
	w_{0}w_{1}\right\rangle_{\beta=0}}_{-1};
\,\,\,s_j=2,s_0=2,p_1=2,p_3=1,\nonumber
\\
K_{2,1}^{l_1}K_{2,2}^{l_1-j}K_{1,1}^{l_3}K_{1,2}^{l_3-j}
\underbrace{\left\langle w_{1}w_{2j+1}\,
	w_{2j}w_{2j+1}\,
	w_{0}w_{2j}\,
	w_{0}w_{1}\right\rangle_{\beta=0}}_{1};
\,\,\,s_j=2,s_0=2,p_1=2,p_3=2. \label{eq:16linesI}
\end{eqnarray}
Now, we introduce the following notation:
\be
R_1:=\left(\begin{matrix}
	l_1-j\\
	l_1
\end{matrix}\right),
\hspace{1cm}
R_3:=\left(\begin{matrix}
	l_3-j\\
	l_3
\end{matrix}\right),
\hspace{1cm}
S:=\left(\begin{matrix}
	s_j\\
	s_0
\end{matrix}\right),
\ee
the spinor ``inversion":
\be
\tilde{1}:=2,\hspace{1cm}\tilde{2}:=1,
\ee
and the following notation for the vector components
\be
v \equiv  \left(\begin{matrix}
	v(1)\\
	v(2)
\end{matrix}\right).
\ee

This allows us to summarise the above $16$ lines in a compact formula:
\begin{equation}
\left(I\right)_{j\neq0}=\sum_{j\neq0}\sum_{l_1,l_3}\sum_{s_j,s_0=1}^{2}\sum_{p_1,p_3=1}^{2}
K_{S(p_1),1}^{R_1(p_1)}K_{S(\tilde{p_1}),2}^{R_1(\tilde{p_1})}
K_{\tilde{S}(p_3),1}^{R_3(p_3)}K_{\tilde{S}(\tilde{p_3}),2}^{R_3(\tilde{p_3})}\left(-1\right)^{s_j+s_0+p_1+p_3},\label{eq:Ijn0}
\end{equation}
where the summation runs over the $3$ spatial indices: $j$, $l_1$, $l_3$ and $4$ permutations: $s_j$, $s_0$, $p_1$, $p_3$. Here, $s_j$ denotes the "spin" of the $j$-type fermion in the first pair of slots ($w_{2j+s_j-1}$) and $s_0$ the "spin" of the $0$-type fermion in the first two slots ($w_{s_0-1}$). Furthermore, $p_1$ denotes the permutation of fermions in the first pair of slots and $p_3$ the permutation of fermions in the third pair of slots. Note also that
\be
\left(II\right)_{j\neq0}=-\left(I\right)_{j\neq0}.
\ee

\textbf{b) $j=0$} 
\newline

\noindent In this case, we have to fill in the following correlator:
\begin{equation}
\left\langle \underline{\,\,\,\,\,\,}\,\, \underline{\,\,\,\,\,\,}w_{0}w_{1}\underline{\,\,\,\,\,\,}\,\, \underline{\,\,\,\,\,\,}w_{0}w_{1}\right\rangle_{\beta=0}.
\end{equation}
Since the present fermions are already contracted, we can fill the empty slots with arbitrary two pairs of fermions $w_{2\bar{j}+s_j-1}$, $w_{2\bar{j}+s_j-1}$ and $w_{2l+s_0-1}$,$w_{2l+s_0-1}$. Again, we want to have
\be
\left\langle w_{0}w_{1}\underline{\,\,\,\,\,\,}\,\, \underline{\,\,\,\,\,\,}\,\,\underline{\,\,\,\,\,\,}\,\, \underline{\,\,\,\,\,\,}w_{0}w_{1}\right\rangle_{\beta=0}=-\left\langle \underline{\,\,\,\,\,\,}\,\, \underline{\,\,\,\,\,\,}w_{0}w_{1}\underline{\,\,\,\,\,\,}\,\, \underline{\,\,\,\,\,\,}w_{0}w_{1}\right\rangle_{\beta=0},
\ee
so that the terms in \eqref{eq:OTOCmagZ} do not end up cancelling out. It is easy to check that the only way to achieve this is to set either $\bar{j}$ or $l$ to zero, with the remaining index being non-zero. We choose $l=0$ and $\bar{j}\neq0$. As before, we again have $16$ possible permutations:
\begin{eqnarray}
K_{1,1}^{l_1-\bar{j}}K_{1,2}^{l_1}K_{1,1}^{l_3-\bar{j}}K_{1,2}^{l_3}
\underbrace{\left\langle w_{2j}w_{0}\,
	w_{0}w_{1}\,
	w_{2\bar{j}}w_{0}\,
	w_{0}w_{1}\right\rangle_{\beta=0}}_{-1};
\,\,\,s_{\bar{j}}=1,s_0=1,p_1=1,p_3=1,\nonumber
\\
K_{1,1}^{l_1-\bar{j}}K_{1,2}^{l_1}K_{1,1}^{l_3}K_{1,2}^{l_3-\bar{j}}
\underbrace{\left\langle w_{2\bar{j}}w_{0}\,
	w_{0}w_{1}\,
	w_{0}w_{2\bar{j}}\,
	w_{0}w_{1}\right\rangle_{\beta=0}}_{1};
\,\,\,s_{\bar{j}}=1,s_0=1,p_1=1,p_3=2,\nonumber
\\
K_{1,1}^{l_1}K_{1,2}^{l_1-\bar{j}}K_{1,1}^{l_3-\bar{j}}K_{1,2}^{l_3}
\underbrace{\left\langle w_{0}w_{2\bar{j}}\,
	w_{0}w_{1}\,
	w_{2\bar{j}}w_{0}\,
	w_{0}w_{1}\right\rangle_{\beta=0}}_{1};
\,\,\,s_{\bar{j}}=1,s_0=1,p_1=2,p_3=1,\nonumber
\\
K_{1,1}^{l_1}K_{1,2}^{l_1-\bar{j}}K_{1,1}^{l_3}K_{1,2}^{l_3-\bar{j}}
\underbrace{\left\langle w_{0}w_{2\bar{j}}\,
	w_{0}w_{1}\,
	w_{0}w_{2\bar{j}}\,
	w_{0}w_{1}\right\rangle_{\beta=0}}_{-1};
\,\,\,s_{\bar{j}}=1,s_0=1,p_1=2,p_3=2,\nonumber
\end{eqnarray}

\begin{eqnarray}
K_{1,1}^{l_1-\bar{j}}K_{2,2}^{l_1}K_{1,1}^{l_3-\bar{j}}K_{2,2}^{l_3}
\underbrace{\left\langle w_{2\bar{j}}w_{1}\,
	w_{0}w_{1}\,
	w_{2\bar{j}}w_{1}\,
	w_{0}w_{1}\right\rangle_{\beta=0}}_{-1};
\,\,\,s_{\bar{j}}=1,s_0=2,p_1=1,p_3=1,\nonumber
\\
K_{1,1}^{l_1-\bar{j}}K_{2,2}^{l_1}K_{2,1}^{l_3}K_{1,2}^{l_3-\bar{j}}
\underbrace{\left\langle w_{2\bar{j}}w_{1}\,
	w_{0}w_{1}\,
	w_{1}w_{2\bar{j}}\,
	w_{0}w_{1}\right\rangle_{\beta=0}}_{1};
\,\,\,s_{\bar{j}}=1,s_0=2,p_1=1,p_3=2,\nonumber
\\
K_{2,1}^{l_1}K_{1,2}^{l_1-\bar{j}}K_{1,1}^{l_3-\bar{j}}K_{2,2}^{l_3}
\underbrace{\left\langle w_{1}w_{2\bar{j}}\,
	w_{0}w_{1}\,
	w_{2\bar{j}}w_{1}\,
	w_{0}w_{1}\right\rangle_{\beta=0}}_{1};
\,\,\,s_{\bar{j}}=1,s_0=2,p_1=2,p_3=1,\nonumber
\\
K_{2,1}^{l_1}K_{1,2}^{l_1-\bar{j}}K_{2,1}^{l_3}K_{1,2}^{l_3-\bar{j}}
\underbrace{\left\langle w_{1}w_{2\bar{j}}\,
	w_{0}w_{1}\,
	w_{1}w_{2\bar{j}}\,
	w_{0}w_{1}\right\rangle_{\beta=0}}_{-1};
\,\,\,s_{\bar{j}}=1,s_0=2,p_1=2,p_3=2,\nonumber
\end{eqnarray}

\begin{eqnarray}
K_{2,1}^{l_1-\bar{j}}K_{1,2}^{l_1}K_{2,1}^{l_3-\bar{j}}K_{1,2}^{l_3}
\underbrace{\left\langle w_{2\bar{j}+1}w_{0}\,
	w_{0}w_{1}\,
	w_{2\bar{j}+1}w_{0}\,
	w_{0}w_{1}\right\rangle_{\beta=0}}_{-1};
\,\,\,s_{\bar{j}}=2,s_0=1,p_1=1,p_3=1,\nonumber
\\
K_{2,1}^{l_1-\bar{j}}K_{1,2}^{l_1}K_{1,1}^{l_3}K_{2,2}^{l_3-\bar{j}}
\underbrace{\left\langle w_{2\bar{j}+1}w_{0}\,
	w_{0}w_{1}\,
	w_{0}w_{2\bar{j}+1}\,
	w_{0}w_{1}\right\rangle_{\beta=0}}_{1};
\,\,\,s_{\bar{j}}=2,s_0=1,p_1=1,p_3=2,\nonumber
\\
K_{1,1}^{l_1}K_{2,2}^{l_1-\bar{j}}K_{2,1}^{l_3-\bar{j}}K_{1,2}^{l_3}
\underbrace{\left\langle w_{0}w_{2\bar{j}+1}\,
	w_{0}w_{1}\,
	w_{2\bar{j}+1}w_{0}\,
	w_{0}w_{1}\right\rangle_{\beta=0}}_{1};
\,\,\,s_{\bar{j}}=2,s_0=1,p_1=2,p_3=1,\nonumber
\\
K_{1,1}^{l_1}K_{2,2}^{l_1-\bar{j}}K_{1,1}^{l_3}K_{2,2}^{l_3-\bar{j}}
\underbrace{\left\langle w_{0}w_{2\bar{j}+1}\,
	w_{0}w_{1}\,
	w_{0}w_{2\bar{j}+1}\,
	w_{0}w_{1}\right\rangle_{\beta=0}}_{-1};
\,\,\,s_{\bar{j}}=2,s_0=1,p_1=2,p_3=2,\nonumber
\end{eqnarray}

\begin{eqnarray}
K_{2,1}^{l_1-\bar{j}}K_{2,2}^{l_1}K_{2,1}^{l_3-\bar{j}}K_{2,2}^{l_3}
\underbrace{\left\langle w_{2\bar{j}+1}w_{1}\,
	w_{0}w_{1}\,
	w_{2\bar{j}+1}w_{1}\,
	w_{0}w_{1}\right\rangle_{\beta=0}}_{-1};
\,\,\,s_{\bar{j}}=2,s_0=2,p_1=1,p_3=1,\nonumber
\\
K_{2,1}^{l_1-\bar{j}}K_{2,2}^{l_1}K_{2,1}^{l_3}K_{2,2}^{l_3-\bar{j}}
\underbrace{\left\langle w_{2\bar{j}+1}w_{1}\,
	w_{0}w_{1}\,
	w_{1}w_{2\bar{j}+1}\,
	w_{0}w_{1}\right\rangle_{\beta=0}}_{1};
\,\,\,s_{\bar{j}}=2,s_0=2,p_1=1,p_3=2,\nonumber
\\
K_{2,1}^{l_1}K_{2,2}^{l_1-\bar{j}}K_{2,1}^{l_3-\bar{j}}K_{2,2}^{l_3}
\underbrace{\left\langle w_{1}w_{2\bar{j}+1}\,
	w_{0}w_{1}\,
	w_{2\bar{j}+1}w_{1}\,
	w_{0}w_{1}\right\rangle_{\beta=0}}_{1};
\,\,\,s_{\bar{j}}=2,s_0=2,p_1=2,p_3=1,\nonumber
\\
K_{2,1}^{l_1}K_{2,2}^{l_1-\bar{j}}K_{2,1}^{l_3}K_{2,2}^{l_3-\bar{j}}
\underbrace{\left\langle w_{1}w_{2\bar{j}+1}\,
	w_{0}w_{1}\,
	w_{1}w_{2\bar{j}+1}\,
	w_{0}w_{1}\right\rangle_{\beta=0}}_{-1};
\,\,\,s_{\bar{j}}=2,s_0=2,p_1=2,p_3=2,
\end{eqnarray}
which we can express as
\begin{equation}
\left(I\right)_{j=0}=\sum_{\bar{j}\neq0}\sum_{l_1,l_3}\sum_{s_{\bar{j}},s_0=1}^{2}\sum_{p_1,p_3=1}^{2}
K_{S(p_1),1}^{R_1(p_1)}K_{S(\tilde{p_1}),2}^{R_1(\tilde{p_1})}
K_{S(p_3),1}^{R_3(p_3)}K_{S(\tilde{p_3}),2}^{R_3(\tilde{p_3})}\left(-1\right)^{p_1+p_3}\left(-1\right).\label{eq:Ij0}
\end{equation}
where we used $R_i:=\left(\begin{matrix}
l_i-\bar{j}\\
l_i
\end{matrix}\right)$ and $S:=\left(\begin{matrix}
s_{\bar{j}}\\
s_0
\end{matrix}\right)$. 

Renaming the dummy summation index in \eqref{eq:Ij0} from $\bar{j}$ to $j$, summing \eqref{eq:Ij0} and \eqref{eq:Ijn0} and plugging it together with $\left(I\right)=-\left(II\right)$ into the expression \eqref{eq:OTOCmagZ} for $c_z(t)$, we finally recover Eq. \eqref{eq:OTOCsimplified} from the main text:
\begin{eqnarray}
c_z(t)&=&-4\sum_{j\neq0}\sum_{l_1,l_3}\sum_{s_j,s_0=1}^{2}\sum_{p_1,p_3=1}^{2}
\left(-1\right)^{p_1+p_3}K_{S(p_1),1}^{R_1(p_1)}(t)K_{S(\tilde{p_1}),2}^{R_1(\tilde{p_1})}(t)\cdot\nonumber\\
&&
\hspace{1.5cm}\cdot\left[\left(-1\right)^{s_j+s_0}K_{\tilde{S}(p_3),1}^{R_3(p_3)}(t)K_{\tilde{S}(\tilde{p_3}),2}^{R_3(\tilde{p_3})}(t) - K_{S(p_3),1}^{R_3(p_3)}(t)K_{S(\tilde{p_3}),2}^{R_3(\tilde{p_3})}(t)\right].\label{eq:OTOCsimplifiedApp}
\end{eqnarray}
Eq. \eqref{eq:OTOCsimplifiedApp} is a general expression for the dOTOC of transverse magnetisation and holds for any free fermion model.

\section{Application of the formula for the dOTOC of $M_z$}\label{app:ApplicationFormulaOTOC}
The equation \eqref{eq:OTOCsimplifiedApp} can be used in two ways: for exact numerical computation at intermediate times and for the analytical computation of the long-time asymptotics.

\subsection{Intermediate times}\label{app:NumaricalApplication}
For intermediate times, $t<50$, we proceed by first computing the power $\mathcal{U}^t$ of KI Floquet propagator \eqref{eq:KIFloquetFourierApp} for given numerical values of the parameters $J$ and $h$:
\begin{equation}
\mathcal{U}^t(J,h,\theta)=\sum_{n=-t}^{t}U_n(J,h)e^{\ii n\theta},\label{eq:PowersU}
\end{equation}
where $U_n$ are $2\times2$ matrices whose elements depend only on $J$ and $h$. It is then easy to see from \eqref{eq:SymbolFinalApp} that
\begin{equation}
K^l(t)=\frac{1}{2\pi}\int_{-\pi}^{\pi}\dd\theta e^{-\ii \theta l} \mathcal{U}^t(\theta)=U_l, 
\end{equation}
thereby enabling a direct computation of the real space propagator for given numerical values of $J$ and $h$.

From the above calculation, we also learn that $K^l(t)\neq0$ only for $\left|l\right|\leq t$. This is a direct observation of the fact that the information in KI spreads in a sharp causal-cone with the speed of propagation equal to $1$. Hereon, it follows that the sums over $j$, $l_1$ and $l_3$ in \eqref{eq:OTOCsimplifiedApp} do not need to be taken over the entire $\ZZ$ but only over the finite intervals $j\in\left[-2t,2t\right]-\left\{0\right\}$, $l_1,l_3\in\left[\max\left(-t,j-t\right),\min\left(t,j+t\right)\right]$. The number of terms in this sum is proportional to $t^3$ so the summation can be efficiently carried out for intermediate $t$. The result is numerically exact.

\subsection{Long-time asymptotics}\label{app:Asymptotics}
To find the long-time asymptotic behavior of $c_z(t)$, we express \eqref{eq:OTOCsimplifiedApp} using \eqref{eq:SymbolFinalApp}:
\begin{eqnarray}
c_z(t)&=&-4\sum_{j\neq0}\sum_{l_1,l_3}\sum_{s_j,s_0=1}^{2}\sum_{p_1,p_3=1}^{2}
\left(-1\right)^{p_1+p_3} \left(\frac{1}{2\pi}\right)^4\int_{-\pi}^{\pi}\dd\theta\int_{-\pi}^{\pi}\dd\theta_1\int_{-\pi}^{\pi}\dd\theta_2\int_{-\pi}^{\pi}\dd\theta_3\nonumber\\
&&
\cdot e^{-\ii \theta R_1(p_1)}\left[\mathcal{U}^{t}(\theta)\right]_{S(p_1),1}e^{-\ii \theta_1 R_1(\tilde{p_1})}\left[\mathcal{U}^{t}\left(\theta_1\right)\right]_{S(\tilde{p_1}),2}\cdot\nonumber\\
&&
\hspace{0cm}\cdot\left\{\left(-1\right)^{s_j+s_0}e^{-\ii \theta_2 R_3(p_3)}\left[\mathcal{U}^{t}(\theta_2)\right]_{\tilde{S}(p_3),1}e^{-\ii \theta_3 R_3(\tilde{p_3})}\left[\mathcal{U}^{t}(\theta_3)\right]_{\tilde{S}(\tilde{p_3}),2}\right.-\nonumber\\
&&
\hspace{0cm}-\left.e^{-\ii \theta_2 R_3(p_3)}\left[\mathcal{U}^{t}(\theta_2)\right]_{S(p_3),1}e^{-\ii \theta_3 R_3(\tilde{p_3})}\left[\mathcal{U}^{t}(\theta_3)\right]_{S(\tilde{p_3}),2}\right\}.
\end{eqnarray}
Introducing the following notation
\begin{equation}
\Theta_1=\left(\begin{matrix}
\theta\\
\theta_1
\end{matrix}\right),
\hspace{1cm}
\Theta_3=\left(\begin{matrix}
\theta_2\\
\theta_3
\end{matrix}\right),
\end{equation}
we then have 
\begin{eqnarray}
c_z(t)&=&-4 \,     \sum_{j}\sum_{l_1,l_3}\sum_{s_j,s_0=1}^{2}\sum_{p_1,p_3=1}^{2}
\left(-1\right)^{p_1+p_3} \left(\frac{1}{2\pi}\right)^4\int_{-\pi}^{\pi}\dd\theta\int_{-\pi}^{\pi}\dd\theta_1\int_{-\pi}^{\pi}\dd\theta_2\int_{-\pi}^{\pi}\dd\theta_3\nonumber\\
&&
\hspace{0.5cm}\cdot e^{-\ii (\theta+\theta_1)l_1}e^{-\ii (\theta_2+\theta_3)l_3}e^{\ii \left(\Theta_1(p_1)+\Theta_3(p_3)\right)j} \left[\mathcal{U}^{t}(\theta)\right]_{S(p_1),1}\left[\mathcal{U}^{t}\left(\theta_1\right)\right]_{S(\tilde{p_1}),2}\cdot\nonumber\\
&&
\hspace{1cm}\cdot\left\{\left(-1\right)^{s_j+s_0}\left[\mathcal{U}^{t}(\theta_2)\right]_{\tilde{S}(p_3),1}\left[\mathcal{U}^{t}(\theta_3)\right]_{\tilde{S}(\tilde{p_3}),2} - \left[\mathcal{U}^{t}(\theta_2)\right]_{S(p_3),1}\left[\mathcal{U}^{t}(\theta_3)\right]_{S(\tilde{p_3}),2}\right\}+ \nonumber\\
&&+4 \, \sum_{l_1,l_3}\sum_{s_j,s_0=1}^{2}\sum_{p_1,p_3=1}^{2}
\left(-1\right)^{p_1+p_3}\cdot \left(\frac{1}{2\pi}\right)^4\int_{-\pi}^{\pi}\dd\theta\int_{-\pi}^{\pi}\dd\theta_1\int_{-\pi}^{\pi}\dd\theta_2\int_{-\pi}^{\pi}\dd\theta_3\nonumber\\
&&
\hspace{0.5cm}\cdot e^{-\ii (\theta+\theta_1)l_1}e^{-\ii (\theta_2+\theta_3)l_3} \left[\mathcal{U}^{t}(\theta)\right]_{S(p_1),1}\left[\mathcal{U}^{t}\left(\theta_1\right)\right]_{S(\tilde{p_1}),2}\cdot\nonumber\\
&&
\hspace{1cm}\cdot\left\{\left(-1\right)^{s_j+s_0}\left[\mathcal{U}^{t}(\theta_2)\right]_{\tilde{S}(p_3),1}\left[\mathcal{U}^{t}(\theta_3)\right]_{\tilde{S}(\tilde{p_3}),2}- \left[\mathcal{U}^{t}(\theta_2)\right]_{S(p_3),1}\left[\mathcal{U}^{t}(\theta_3)\right]_{S(\tilde{p_3}),2}\right\}.
\end{eqnarray}
Here, we took the $j$ sum over the entire $\ZZ$ in the first term and then subtracted the $j=0$ case in the second term. Performing (formally) the $j$, $l_1$ and $l_3$ summations and taking into account $\sum_n e^{\ii n x}=2\pi\sum_k\delta\left(x-k2\pi\right)$, we get
\begin{eqnarray}
c_z(t)&=&-4  \sum_{s_j,s_0=1}^{2}\sum_{p_1,p_3=1}^{2}
\left(-1\right)^{p_1+p_3}\cdot\nonumber\\
&&
\cdot\Biggl[\frac{1}{2\pi}\int_{-\pi}^{\pi}\dd\theta\int_{-\pi}^{\pi}\dd\theta_1\int_{-\pi}^{\pi}\dd\theta_2\int_{-\pi}^{\pi}\dd\theta_3\nonumber\\
&&
\hspace{0.5cm}\cdot \delta(\theta+\theta_1)\delta(\theta_2+\theta_3)\delta\left(\Theta_1(p_1)+\Theta_3(p_3)\right)\left[\mathcal{U}^{t}(\theta)\right]_{S(p_1),1}\left[\mathcal{U}^{t}\left(\theta_1\right)\right]_{S(\tilde{p_1}),2}\cdot\nonumber\\
&&
\hspace{1cm}\cdot\left\{\left(-1\right)^{s_j+s_0}\left[\mathcal{U}^{t}(\theta_2)\right]_{\tilde{S}(p_3),1}\left[\mathcal{U}^{t}(\theta_3)\right]_{\tilde{S}(\tilde{p_3}),2}- \left[\mathcal{U}^{t}(\theta_2)\right]_{S(p_3),1}\left[\mathcal{U}^{t}(\theta_3)\right]_{S(\tilde{p_3}),2}\right\}- \nonumber \\
&&-\left(\frac{1}{2\pi}\right)^2\int_{-\pi}^{\pi}\dd\theta\int_{-\pi}^{\pi}\dd\theta_1\int_{-\pi}^{\pi}\dd\theta_2\int_{-\pi}^{\pi}\dd\theta_3\nonumber\\
&&
\hspace{0.5cm}\cdot \delta(\theta+\theta_1)\delta(\theta_2+\theta_3) \left[\mathcal{U}^{t}(\theta)\right]_{S(p_1),1}\left[\mathcal{U}^{t}(\theta_1)\right]_{S(\tilde{p_1}),2}\cdot\nonumber\\
&&
\hspace{1cm}\cdot\left\{\left(-1\right)^{s_j+s_0}\left[\mathcal{U}^{t}(\theta_2)\right]_{\tilde{S}(p_3),1}\left[\mathcal{U}^{t}(\theta_3)\right]_{\tilde{S}(\tilde{p_3}),2}- \left[\mathcal{U}^{t}(\theta_2)\right]_{S(p_3),1}\left[\mathcal{U}^{t}(\theta_3)\right]_{S(\tilde{p_3}),2}\right\}\Biggr].
\end{eqnarray}
Finally, integrating over $\theta_1$, $\theta_2$ and $\theta_3$ in the first term and $\theta_1$ and $\theta_3$ in the second, we get:
\begin{eqnarray}
c_z(t)&=&-4\sum_{s_j,s_0=1}^{2}\sum_{p_1,p_3=1}^{2}
\left(-1\right)^{p_1+p_3}\nonumber\\
&&
\cdot\Biggl[\frac{1}{2\pi}\int_{-\pi}^{\pi}\dd\theta \left[\mathcal{U}^{t}(\theta)\right]_{S(p_1),1}\left[\mathcal{U}^{t}(-\theta)\right]_{S(\tilde{p_1}),2}\cdot\nonumber\\
&&
\cdot\left\{\left(-1\right)^{s_j+s_0}\left[\mathcal{U}^{t}\left(-\left(-1\right)^{p1+p3}\theta\right)\right]_{\tilde{S}(p_3),1}\left[\mathcal{U}^{t}\left(\left(-1\right)^{p1+p3}\theta\right)\right]_{\tilde{S}(\tilde{p_3}),2}\right.-\nonumber\\
&&
\hspace{0cm}-\left.\left[\mathcal{U}^{t}\left(-\left(-1\right)^{p1+p3}\theta\right)\right]_{S(p_3),1}\left[\mathcal{U}^{t}\left(\left(-1\right)^{p1+p3}\theta\right)\right]_{S(\tilde{p_3}),2}\right\}- \nonumber\\
&&-\left(\frac{1}{2\pi}\right)^2\int_{-\pi}^{\pi}\dd\theta\left[\mathcal{U}^{t}(\theta)\right]_{S(p_1),1}\left[\mathcal{U}^{t}(-\theta)\right]_{S(\tilde{p_1}),2}\cdot\nonumber\\
&&
\hspace{1cm}\cdot\int_{-\pi}^{\pi}\dd\theta_2\left\{\left(-1\right)^{s_j+s_0}\left[\mathcal{U}^{t}(\theta_2)\right]_{\tilde{S}(p_3),1}\left[\mathcal{U}^{t}(-\theta_2)\right]_{\tilde{S}(\tilde{p_3}),2}\right.-\nonumber\\
&&
\hspace{2cm}-\left.\left[\mathcal{U}^{t}(\theta_2)\right]_{S(p_3),1}\left[\mathcal{U}^{t}(-\theta_2)\right]_{S(\tilde{p_3}),2}\right\}\Biggr]\\
&=:&\int_{-\pi}^{\pi}\dd\theta \, I_1\left(t,\theta\right)+\int_{-\pi}^{\pi}\dd\theta \int_{-\pi}^{\pi}\dd\theta_2\, I_2\left(t,\theta,\theta_2\right).\label{eq:IntegralTheta}
\end{eqnarray}
It can be shown that the integrand $I_2$ vanishes:
\be
I_2\left(t,\theta,\theta_2\right)=0.
\ee
The remaining integration over $\theta$ in $I_1$ is in general difficult to perform but we can find the asymptotic behavior for large $t$.

To complete this task we can take advantage of the fact that $\mathcal{U}$ is a unitary matrix. Using the form \eqref{eq:FloquetUnitary}, the powers of $\mathcal{U}$ are simply:
\begin{equation}
\mathcal{U}^t(\theta)=V^\dagger(\theta) \left(\begin{matrix}
e^{\ii t \kappa(\theta)} && \\
&& e^{\ii t \lambda(\theta)}
\end{matrix}\right) V(\theta).\label{eq:PowerU}
\end{equation}

We can then compute the integral \eqref{eq:IntegralTheta} in the large $t$ regime by using the  stationary phase approximation \cite{SircaHorvat2012}:
\begin{equation}
\int_{a}^{b}\dd\theta \phi (\theta) e^{\ii t \psi(\theta)}\sim \sum_j \phi(\xi_j) \sqrt{\frac{2\pi}{t\left|\psi''(\xi_j)\right|}}\exp\left\{\ii \left[t\psi(\xi_j)+\frac{\pi}{4}\text{sign}\left(\psi''(\xi_j)\right)\right]\right\},\label{eq:StationaryPhase}
\end{equation}
where $\xi_j$ denotes (all of) the local extrema of $\psi(\theta)$, i.e. $\psi'(\xi_j)=0$, on the interval $\left[a,b\right]$. 

\subsubsection{The case of the kicked quantum Ising model}

The considerations so far have been general and can be applied to any quadratic fermion model. For the KI model, we can use the expressions for eigenvalues and eigenvectors from Section \ref{app:SpectrumKI}. 

Let us introduce the following notation:
\begin{eqnarray}
\left\{V\right\}(\theta)&:=&\left\{v_{11}(\theta),v_{12}(\theta),v_{21}(\theta),v_{22}(\theta),\right.\nonumber\\
&&\overline{v_{11}}(\theta),\overline{v_{12}}(\theta),\overline{v_{21}}(\theta),\overline{v_{22}}(\theta),\nonumber\\
&&v_{11}(-\theta),v_{12}(-\theta),v_{21}(-\theta),v_{22}(-\theta),\nonumber\\
&&\left.\overline{v_{11}}(-\theta),\overline{v_{12}}(-\theta),\overline{v_{21}}(-\theta),\overline{v_{22}}(-\theta)\right\}.
\end{eqnarray}
Plugging the results from Section \eqref{app:SpectrumKI} into \eqref{eq:IntegralTheta}, we see that integrand $I_1$ can be written the following form:
\begin{eqnarray}
I_1\left(t,\theta\right)&=&P_0\left[\left\{V\right\}(\theta)\right]+\nonumber\\
&&+P_{-2}\left[\left\{V\right\}(\theta)\right]e^{-2\ii t\kappa(\theta)}+P_{2}\left[\left\{V\right\}(\theta)\right]e^{2\ii t\kappa(\theta)}+\nonumber\\
&&+P_{-4}\left[\left\{V\right\}(\theta)\right]e^{-4\ii t\kappa(\theta)}+P_{4}\left[\left\{V\right\}(\theta)\right]e^{4\ii t\kappa(\theta)}.\label{eq:IntegrandKI1}
\end{eqnarray}
Here, all $P$'s are polynomials in their arguments. The indices denote the power of the term $e^{\ii t\kappa(\theta)}$ multiplying a particular polynomial. All eigenvactor components $\left\{V\right\}$ and the eigenvalue $\kappa$ also depend on parameters $J$ and $h$. 

The large $t$ behavior of the integrals $\int_{-\pi}^{\pi}\dd\theta$ of all the terms except for $P_0$ can be obtained using stationary phase approximation \eqref{eq:StationaryPhase}. Plugging in the elements of the eigenvectors at stationary points of $\kappa(\theta)$, that is \eqref{eq:StationaryVector}, we see that $P_{-2}$, $P_{2}$, $P_{-4}$, $P_{4}$ vanish at these points. This means that in the large $t$ regime, $c_z(t)$ is constant for the transverse field KI. 

The only remaining integral
\be
\lim_{t\rightarrow\infty}c_z(t)=\int_{-\pi}^{\pi}\dd\theta\, P_0\left[\left\{V\right\}(J,h,\theta)\right]
\ee
can be evaluated numerically to get the asymptotic (constant) value of the $c_z(t)$ for any given $J$ and $h$ (See the inset of Figure 1 of the main text). 

\end{document}